# Enabling Ultra-Fast Cardiovascular Imaging Across Heterogeneous Clinical Environments with A Generalist Foundation Model and Multimodal Database

Zi Wang[1], Mingkai Huang[2], Zhang Shi[3], Hongjie Hu[4], Lan Lan[5], Hui Zhang[6], Yan Li[7], Xi Hu[4], Qing Lu[8], Zongming Zhu[9], Qiong Yao[10], Yuxiang Dai[6,11], Fanwen Wang[1,12], Yinzhe Wu[1,12], Jun Lyu[13], Qianqian Gao[9], Guangming Xu[9], Zhenxuan Zhang[1], Haosen Zhang[14], Qing Li[14], Guangming Wang[14], Tianxing He[14], Lizhen Lan[14], Siyue Li[15], Le Xue[16], Mengting Sun[14], Yuntong Lyu[17], Junpu Hu[18], Jiayu Zhu[19], Rizwan Ahmad[20,21], Zhengyu Bu[20], Xianling Qian[3], Guanke Cai[10], Ruiyu Cao[6], Weirui Cai[6], Chang Xu[6], Yuyang Ren[22], Feidan Yu[4], Siying Ma[4], Ziqiang Xu[23], Xinran Chen[1], Sha Hua[24], Daniel Kim[25], Yajing Zhang[26], Chen Ouyang[27], Wenjia Bai[28], Jing Qin[29], Yucheng Yang[6], Daniel Rueckert[30], He Wang[6], Qian Tao[31], Claudia Prieto[32,33], Michael Markl[25], Alistair Young[33], Lianming Wu[34], Shuo Wang[35], Chen Qin[36], Mengsu Zeng[3], Xihong Hu[10], Haibo Xu[5], Xiaobo Qu[2,37], Hao Li[6], Guang Yang[1,38,12,33], Chengyan Wang[14]

Multimodal cardiovascular magnetic resonance (CMR) imaging provides comprehensive and non-invasive insights into cardiovascular disease (CVD) diagnosis and underlying mechanisms. Despite decades of advancements, its widespread clinical adoption remains constrained by prolonged scan times, inconsistent image quality, and heterogeneity across medical environments. This underscores the urgent need for a generalist reconstruction foundation model for ultra-fast CMR imaging—one formulated for physics-constrained inverse problems in the sensor (k-space) domain, capable of adapting across diverse imaging scenarios and serving as the essential substrate for all downstream analyses. To enable this goal, we curate MMCMR-427K, the largest and most comprehensive multimodal CMR k-space database to date, comprising 427,465 multi-coil k-space data paired with structured metadata across 13 international centers, 12 CMR modalities, 15 scanners spanning four field strengths, and 17 CVD categories in populations across three continents. Building on this unprecedented resource, we introduce CardioMM, a generalist reconstruction foundation model capable of dynamically adapting to heterogeneous fast CMR imaging scenarios. CardioMM unifies semantic contextual understanding with physics-informed data consistency to deliver robust reconstructions across varied scanners, protocols, and patient presentations. Comprehensive evaluations demonstrate that CardioMM achieves state-of-the-art performance across internal centers and exhibits strong zero-shot generalization to unseen external settings. Importantly, CardioMM supports acceleration up to 24×, providing the first evidence that such extreme acquisition speed can preserve key cardiac phenotypes, quantitative myocardial biomarkers, and diagnostic image quality without compromising clinical integrity. Beyond speed, it also enhances imaging robustness in challenging cardiovascular cases by shortening the vulnerable window for motion and physiological instability. Together, our open-access MMCMR-427K database and CardioMM framework establish a scalable pathway toward high-throughput, high-quality, and clinically accessible multimodal CMR imaging. Beyond cardiovascular imaging, this work underscores the broader clinical promise of foundation modeling for physics-constrained inverse problems in medical imaging systems.

[1]Department of Bioengineering and Imperial-X, Imperial College London, UK. [2]School of Electronic Science and Engineering (National Model Microelectronics College), Xiamen University-Neusoft Medical Magnetic Resonance Imaging Joint Research and Development Center, Fujian Provincial Key Laboratory of Plasma and Magnetic Resonance, Xiamen University, China. [3]Department of Radiology, Zhongshan Hospital, Fudan University, China. [4]Department of Radiology, Sir Run Run Shaw Hospital (SRRSH), Zhejiang University School of Medicine, China. [5]Department of Radiology, Zhongnan Hospital of Wuhan University, China. [6]Institute of Science and Technology for Brain-Inspired Intelligence, Fudan University, China. [7]Department of Radiology, Ruijin Hospital, Shanghai Jiaotong University School of Medicine, China. [8]Department of Radiology, Shanghai East Hospital, Tongji University School of Medicine, China. [9]Department of Radiology, The Affiliated Wuxi People's Hospital of Nanjing Medical University, Wuxi People's Hospital, Wuxi Medical Center, Nanjing Medical University, China. [10]Department of Radiology, Children's hospital of Fudan University, China. [11]Centre for Population Neuroscience and Stratified Medicine (PONS), Department of Psychiatry and Neuroscience, Charité-Universitätsmedizin Berlin, Germany. [12]Cardiovascular Research Centre, Royal Brompton Hospital, UK. [13]School of Computer and Control Engineering, Yantai University, China. [14]Human Phenome Institute and Shanghai Pudong Hospital, Fudan University, China. [15]Hong Kong Centre for Cerebro-cardiovascular Health Engineering, China. [16]Department of Nuclear Medicine/PET Center, Huashan Hospital, Fudan University, China. [17]School of Clinical Medicine, Zhongshan Hospital, Shanghai Medical College, Fudan University, China. [18]Division of Pediatric Cardiology, Department of Pediatrics, The University of Texas Southwestern Medical Center, USA. [19]Collaborative Innovation Department, United Imaging Healthcare Group Co., Ltd., China. [20]Department of Biomedical Engineering, The Ohio State University, USA. [21]Department of Electrical and Computer Engineering, The Ohio State University, USA. [22]School of Biomedical Engineering, ShanghaiTech University, China. [23]Shanghai Fuying Medical Technology Co., Ltd., China. [24]Department of Cardiovascular Medicine, Heart Failure Center, Ruijin Hospital Lu Wan Branch, Shanghai Jiao Tong University School of Medicine, China. [25]Department of Radiology, Feinberg School of Medicine, Northwestern University, USA. [26]Science & Technology Organization, GE Healthcare, China. [27]Institute of Biomedical Engineering, Department of Engineering Science, University of Oxford, UK. [28]Department of Computing and Department of Brain Sciences, Imperial College London, UK. [29]School of Nursing, The Hong Kong Polytechnic University, China. [30]School of Computation, Information and Technology, Technische Universität München, Germany. [31]Department of Imaging Physics, Delft University of Technology, Netherlands. [32]School of Engineering and the iHEALTH Millenium Institute, Pontificia Universidad Católica de Chile, Chile. [33]School of Biomedical Engineering and Imaging Sciences, King's College London, UK. [34]Department of Radiology, Ren Ji Hospital, School of Medicine, Shanghai Jiao Tong University, China. [35]Digital Medical Research Center, School of Basic Medical Sciences, Fudan University, China. [36]Department of Electrical and Electronic Engineering and I-X, Imperial College London, UK. [37]Department of Radiology, the First Affiliated Hospital of Xiamen University, School of Medicine, Xiamen University, China. [38]National Heart and Lung Institute, Imperial College London, UK. Contributed equally to this work: Zi Wang, Mingkai Huang, Zhang Shi, Hongjie Hu, Lan Lan, and Hui Zhang. Corresponding authorship: Xiaobo Qu (quxiaobo@xmu.edu.cn), Xihong Hu (huxihong@dudan.edu.cn), Haibo Xu (xuhaibo@whu.edu.cn), Hao Li (h_li@fudan.edu.cn), Guang Yang (g.yang@imperial.ac.uk), and Chengyan Wang (wangcy@fudan.edu.cn). Guang Yang and Chengyan Wang are co-last authors.

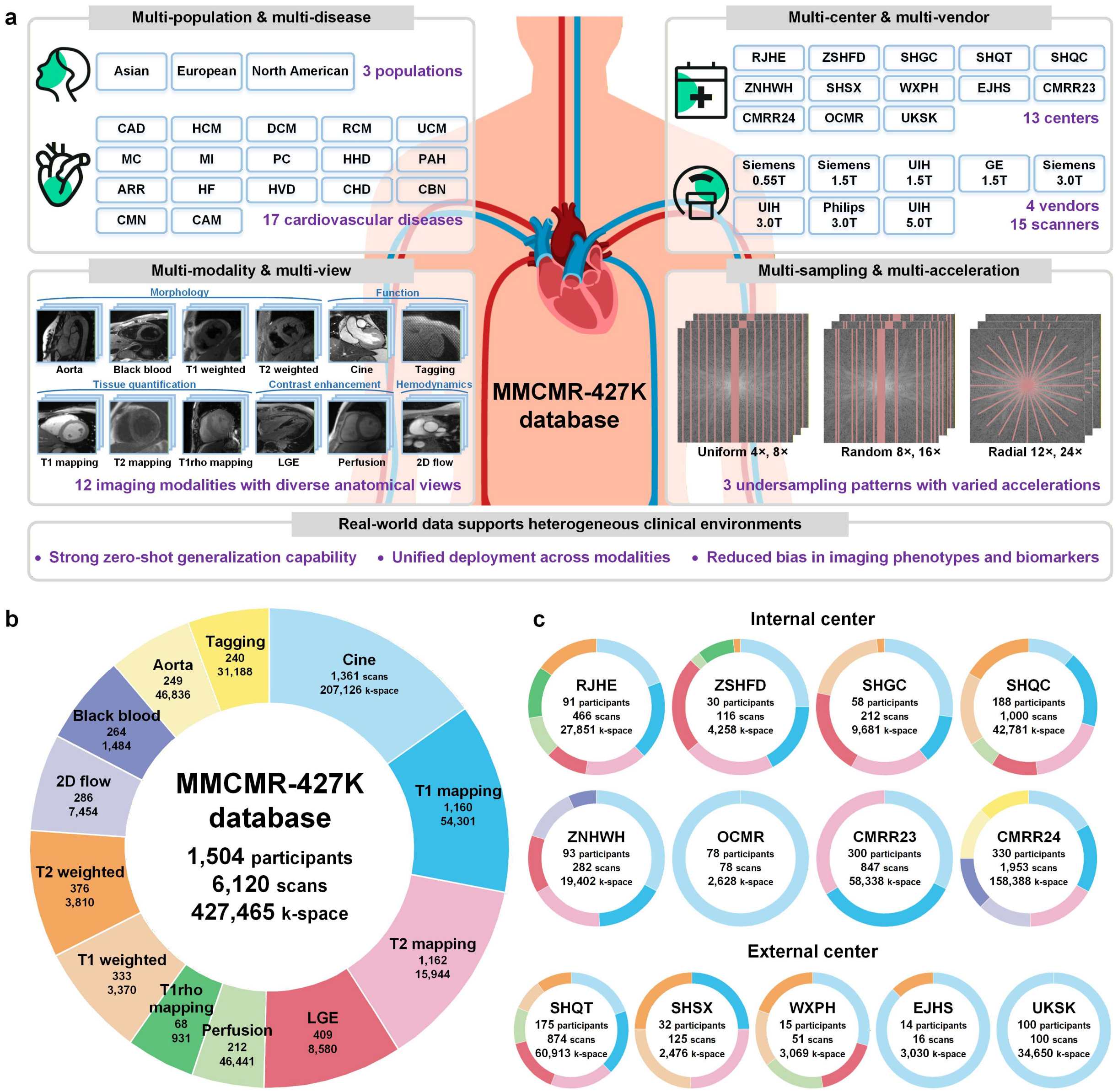

**Fig. 1 | MMCMR-427K, a foundation-scale CMR k-space database spanning populations, diseases, and imaging environments. a**, MMCMR-427K is a large-scale, multi-population, multi-disease, multi-center, multi-vendor, and multimodal CMR k-space database. All cardiovascular diseases are given in abbreviations here, while their full names and detailed information are provided in Supplementary Table 2. **b**, MMCMR-427K comprises 427,465 multi-coil k-space data (approximately 3.5 TB of storage) from 6,120 scans of 1,504 participants. **c**, to facilitate rigorous benchmarking, we categorize 13 worldwide centers into eight internal centers and five external centers. Note: LGE = Late Gadolinium Enhancement. Some vector images are modified from freepik.com and iconfont.cn.

Cardiovascular diseases (CVDs) remain the leading cause of death worldwide and continue to impose a substantial burden on healthcare systems[1-3]. Multimodal cardiovascular magnetic resonance (CMR) imaging, encompassing diverse imaging protocols, provides unparalleled versatility for the comprehensive and non-invasive assessment of cardiac structure, function, perfusion, and tissue characterization. It has become one of the reference standards for CVD diagnosis[4-9]. However, routine CMR examinations are time-consuming (typically 30–60 minutes), forming the principal barrier to integration into time-sensitive clinical workflows[6]. Achieving high-quality multimodal CMR imaging under high accelerations is therefore essential[10-13]. Such capability improves scanner throughput, patient comfort, and resilience to motion artifacts, while facilitating richer multimodal examinations within the fixed time shots to supporting comprehensive clinical decision-making[5,6,14,15].

Accelerated CMR reconstruction constitutes a large-scale, ill-posed inverse problem defined in the sensor domain (i.e., k-space, the raw data of MR scanners in the Fourier domain). Unlike perceptual foundation models that operate in fully observed image space[16-20], reconstruction models must work directly in partially observed sensor space under strict physical constraints. The mapping from undersampled multi-coil k-space data to high-quality images is governed by acquisition geometry, coil sensitivity encoding, and strict data-consistency constraints, rendering the solution space highly structured and physically constrained[10-13,21,22]. In real-world environments, variations in scanner configurations and sampling strategies effectively modify the underlying forward operator; when coupled with high acceleration requirements, these changes further exacerbate the instability of the inverse mapping.

Conventional acceleration techniques such as parallel imaging[10,11] and compressed sensing[12,13] have been developed but remain intrinsically limited in achievable acceleration and clinically viable reconstruction times[15]. Artificial intelligence (AI)-driven approaches offers both higher acceleration in acquisition and reconstruction, yet remains fragile to the substantial heterogeneity of real-world acquisitions, including variations across centers, vendors, protocols, and patient populations[15,21-24]. Such variability fundamentally alters image contrast and sampling characteristics, causing the performance of existing reconstruction methods to degrade or become inconsistent outside their narrow development domains.

In recent years, advances in medical AI[16-21,25,26] have led to the development of generalist foundation models that have achieved impressive performance in post-reconstruction CMR analysis, such as segmentation, classification, and phenotyping[9,27,28]. Nevertheless, most existing efforts focus on a limited set of CMR modalities and presuppose the availability of high-quality images. Yet high-quality images fundamentally depend on reliable and efficient CMR acquisition and reconstruction pipelines. In this context, reliable image reconstruction for fast multimodal CMR imaging (i.e., physics-constrained inverse problems in sensor space), the fundamental prerequisite for downstream analysis, remains at an early stage of investigation[29,30].

A major bottleneck in developing reconstruction foundation models for fast multimodal CMR imaging and subsequent analysis lies in the scale and quality of data. Although several public CMR repositories[31-38] have increased in number over recent years, they are typically fragmented, restricted to specific populations, centers, vendors, CMR modalities, or diseases types, and often lack the raw k-space data and paired metadata required for clinically compatible model training, thereby restricting their usage for real-world reconstruction and analysis tasks. Addressing this gap calls for a large-scale, high-quality, standardized, and multimodal CMR k-space database with paired textual information.

These data limitations cascade into constraints on model design and generalization. Most existing AI-driven CMR image reconstruction models[22,29,30,39] rely exclusively on limited visual information, overlooking rich and clinically meaningful metadata, such as imaging configurations. As a result, their generalization across centers and protocols remains severely constrained, falling short of handling the complexity of CMR in real-world scenarios. A generalist foundation model capable of dynamically adapting to heterogeneous data and fast imaging scenarios is therefore essential to ensure both reconstruction reliability and clinical applicability.

Beyond data and model development, robust validation remains a critical challenge. Most previous studies are confined to single center, a small number of CMR modalities, or evaluations based mainly on conventional image quality metrics, with insufficient emphasis on clinical relevance[6,29,30]. A rigorous and comprehensive evaluation strategy is required, extending beyond visual fidelity to assess diagnostic reliability through key imaging phenotypes and quantitative biomarkers, thereby fostering clinician trust and enabling meaningful clinical translation of AI-driven reconstruction.

In this work, to fill the data gap, we curate MMCMR-427K, the first large-scale, multi-population, multi-disease, multi-center, multi-vendor, and multimodal CMR k-space database (Fig. 1). MMCMR-427K comprises 427,465 multi-coil k-space data from 6,120 scans of 1,504 participants, spanning 13 worldwide centers, 12 CMR modalities, 15 scanners with four field strengths, and 17 CVD categories in populations across three continents. The unified data preparation and quality control pipeline ensures cross-center consistency and reliability. By uniting unprecedented scale, diversity, and paired clinically relevant textual information, MMCMR-427K lays a comprehensive infrastructure for subsequent multimodal CMR reconstruction and analysis.

Based on this resource, we propose CardioMM, a reconstruction foundation model for fast multimodal CMR imaging and analysis (Fig. 2a). CardioMM unrolls the iterative reconstruction process into alternating text-aware image de-aliasing and physics-informed data consistency, thereby incorporating both clinical semantic context and imaging physics. At its core, a text representation module employs a pretrained CLIP text encoder[40] with two learnable projection heads to embed metadata and undersampling texts, enabling dynamic adaptation to diverse imaging scenarios (Supplementary Fig. 2). This design allows CardioMM to maintain broad semantic and imaging knowledge while flexibly adapting to specific tasks, resulting in strong versatility, generalizability and clinical applicability (Figs. 2b-d).

Furthermore, we introduce a comprehensive evaluation strategy that extends beyond conventional image quality metrics to assess broader clinical applicability. By jointly validating image fidelity, imaging phenotype and biomarker reliability, and radiologist judgment, we clearly address key concerns from both engineering and clinical perspectives. In internal scenarios, CardioMM provides state-of-the-art reconstructions across centers and modalities. In external scenarios, CardioMM demonstrates remarkable zero-shot generalization to unseen centers, scanners, and populations, while maintaining robust performance across field strengths from 0.55T to 5.0T.

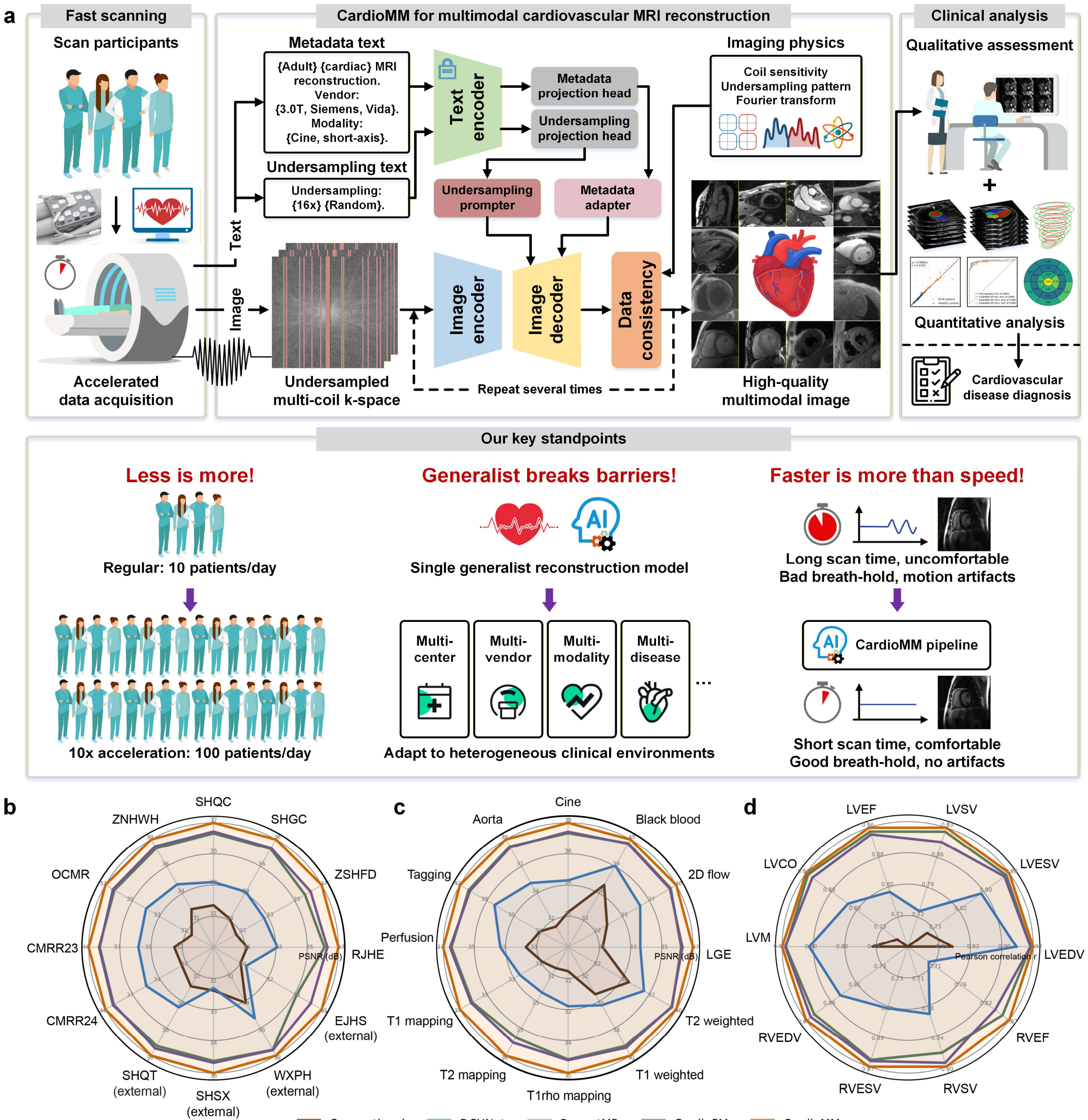

**Fig. 2 | Overview workflow of the proposed CardioMM framework and preliminary results. a**, CardioMM is a generalist reconstruction foundation model for ultra-fast multimodal CMR imaging, which unrolls the iterative reconstruction into alternating text-aware image de-aliasing and physics-informed data consistency, thereby incorporating both clinical semantic context and imaging physics into the reconstruction process. **b-c**, in evaluations across three complementary perspectives, namely cross-center generalization (b), cross-modality generalization (c), and preservation of key imaging phenotypes (d), CardioMM consistently achieves state-of-the-art performance. Note: LVEDV = left ventricular end-diastolic volume, LVESV = left ventricular end-systolic volume, LVSV = left ventricular stroke volume, LVCO = left ventricular cardiac output, LVM = left ventricular mass, LVEF = left ventricular ejection fraction, RVEDV = right ventricular end-diastolic volume, RVESV = right ventricular end-systolic volume, RVSV = right ventricular stroke volume, RVEF = right ventricular ejection fraction. Some vector images are modified from freepik.com.

CardioMM-reconstructed images match the quality of fully sampled references for phenotyping and quantifying cardiovascular myocardial biomarkers, ensuring reliable diagnostic support under high accelerations (8×–24×). In a reader study, CardioMM achieves image quality scores between good and excellent (4.43 out of a 5-point Likert scale), comparable to fully sampled references. The reliability of cardiovascular phenotypes and biomarkers highlights the clinical usefulness of our CardioMM in high-throughput workflows.

In summary, we present a novel database–model–validation synergistic paradigm to advance the full pipeline of multimodal CMR imaging, from ultra-fast acquisition and high-quality reconstruction to clinical meaningful analysis. This study lays the groundwork for integrating reconstruction foundation models into real-world cardiovascular imaging workflows, with strong potential to enable high-throughput and reliable CMR examinations and CVD diagnosis across diverse populations and healthcare environments.

## Results

### MMCMR-427K is a comprehensive CMR k-space database

In this work, we construct MMCMR-427K, the largest and most comprehensive multimodal CMR k-space database to date (Fig. 1a-b). Our MMCMR-427K database contains 427,465 multi-coil k-space data (approximately 3.5 TB) from 6,120 scans of 1,504 participants, covering 17 CVD categories across three populations (Asian, European, and North American). Data were collected from 13 worldwide centers, including four public repositories[31-34] and nine clinical centers, with imaging performed on 15 scanners from four vendors (Siemens, UIH, GE, and Philips) at field strengths ranging from 0.55T to 5.0T. To facilitate rigorous benchmarking, we categorize these centers into internal cohorts (for training, validation, and universal test) and external cohorts (for generalization capability evaluation), enabling systematic assessment across different scenarios (Fig. 1c).

The database spans 12 imaging modalities (e.g., cine, LGE, T1/T2 mapping, perfusion, black blood, tagging) and diverse anatomical views, together with three commonly used undersampling patterns[29,30,41] (uniform, random, radial) at multiple acceleration factors (AFs). This provides a comprehensive testbed for accelerated multimodal CMR image reconstruction and analysis (Fig. 1a). Beyond images, each k-space data is paired with structured scanning metadata (e.g., center, scanner, field strength, imaging protocol), providing semantic information to support the development of text-aware, dynamically adaptive foundation models for generalizable reconstruction across heterogeneous clinical scenarios. More details can be found in Supplementary Note 1.

To ensure consistency and quality, we implemented a unified data preparation pipeline and conducted rigorous quality control procedures, as summaries in Methods. By integrating scale, diversity, and paired metadata, MMCMR-427K represents the most comprehensive, high-quality, and organized CMR k-space database to date, serving as a solid foundation for training and evaluating generalist foundation models in multimodal cardiovascular imaging.

### CardioMM is a CMR reconstruction foundation model

CardioMM is proposed as a generalist reconstruction foundation model for fast multimodal CMR imaging, designed to unify diverse imaging protocols, acquisition settings, and clinical contexts within a single adaptive framework (Fig. 2a). Our model unrolls the iterative reconstruction pipeline into alternating text-aware image de-aliasing modules and physics-informed data consistency modules (See Supplementary Note 2). With this framework, reconstruction is guided simultaneously by clinical semantic contexts and underlying imaging physics, thereby enhancing the reliability and clinical applicability of the reconstructed outcomes.

At the core of CardioMM lies a text representation module that leverages a pretrained CLIP text encoder[40] to embed scan-related descriptions. To ensure robustness and flexibility, we freeze the text encoder to preserve broad semantic knowledge while introducing two learnable projection heads for metadata and undersampling texts, allowing task-specific representations that can be easily extended to additional text types.

On this basis, CardioMM incorporates two complementary mechanisms: the metadata adapter and the undersampling prompter. The metadata adapter injects global semantic context (i.e., patient condition, anatomical region, imaging configuration) into the image decoder, providing both global semantic awareness and adaptive modulation across imaging scenarios. The undersampling prompter captures local artifact priors from undersampling settings (i.e., undersampling pattern, AF), delivering artifact-aware prompts that explicitly inform the network how artifacts manifest under varying undersampling scenarios.

The backbone of the image de-aliasing module is a UNet-like architecture[42] with residual connections and channel attention mechanisms[39,43]. To preserve universal image representations, text information is injected only into the image decoder, allowing the image encoder to remain domain-agnostic while the decoder dynamically adapts its outputs according to semantic and acquisition contexts. By hierarchically combining metadata awareness with undersampling prompts, CardioMM progressively removes aliasing artifacts while maintaining anatomical fidelity (Implementation details are summarized in Methods).

Although the image de-aliasing module relies on explicit priors from metadata and undersampling texts, it remains applicable to unseen combinations of data and text. For unseen scenarios, the text representation module identifies semantically related information closest to the target input and expands it to generate meaningful conditioning (Supplementary Fig. 2). This enables CardioMM to generalize across diverse fast imaging tasks, including those not encountered during training.

By combining semantic awareness with physics-based fidelity, CardioMM acts as a generalizable CMR image reconstruction model that is trained once but can efficiently adapt across diverse fast CMR imaging tasks. Preliminarily, in evaluations across three complementary perspectives, namely cross-center generalization, cross-modality generalization, and preservation of key imaging phenotypes, CardioMM consistently achieves state-of-the-art performance (Figs. 2b-d), highlighting its versatility, generalizability, and potential for real-world cardiovascular imaging.

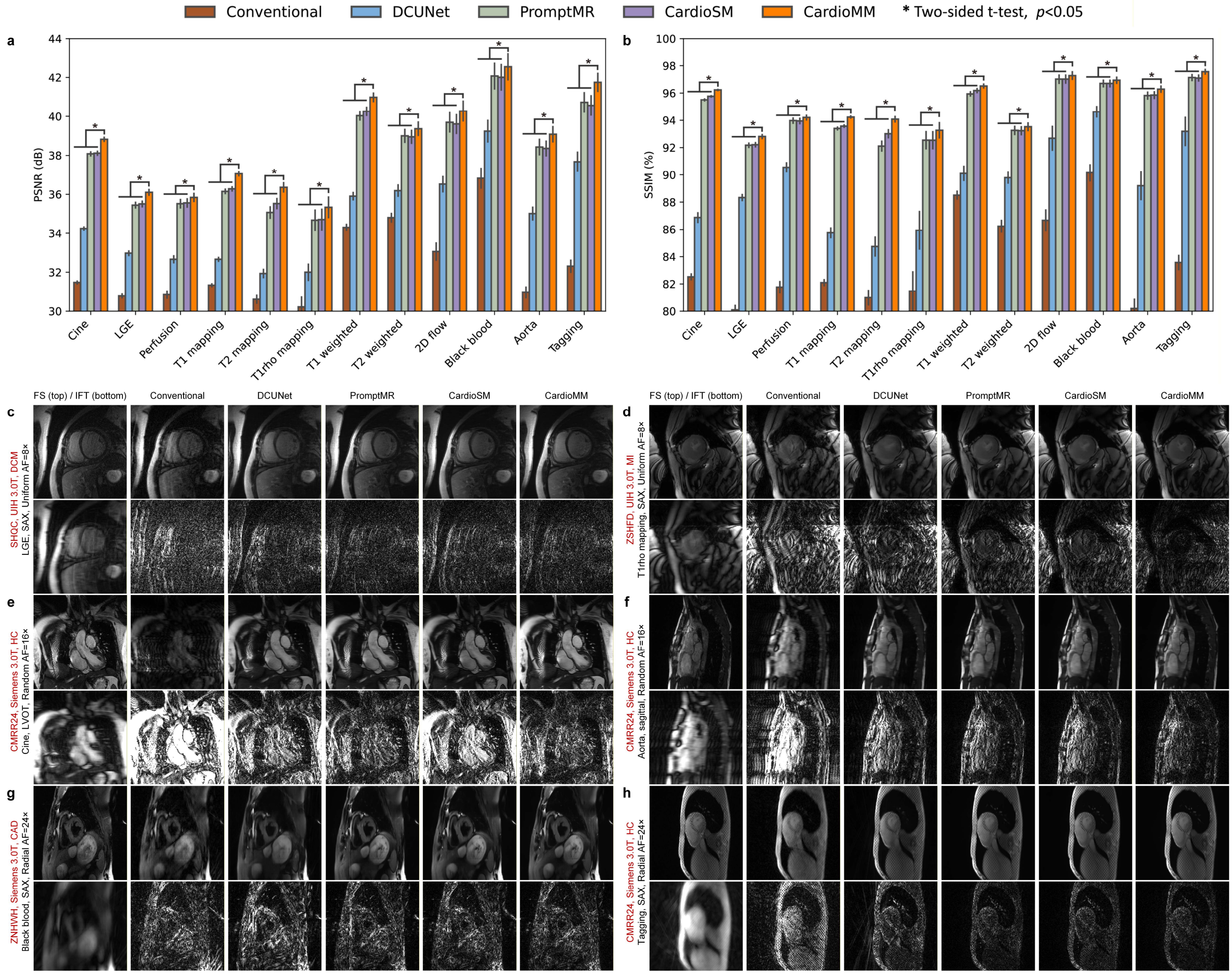

**Fig. 3 | Universal reconstruction across internal scenarios. a-b**, quantitative comparisons of reconstructions are shown for each modality, including PSNR and SSIM. **c-h**, representative reconstruction examples of different methods and their corresponding error maps (scale 0–0.1). Note: This evaluation is conducted across eight internal centers, using three undersampling patterns (uniform, random, radial) with varying AFs (8×–24×). The reported mean values and 95% CIs in the bar charts are computed over all tested data for each modality, respectively. "FS" is the fully sampled reference. "IFT" indicates that using only inverse Fourier transform to reconstruct undersampled k-space leads to images with strong artifacts. CI = confidence interval.

## Rigorous and comprehensive evaluation settings

To comprehensively evaluate the reconstruction and analysis performance of CardioMM, we design a systematic assessment covering both internal and external scenarios.

For the internal scenarios, we first assess universal reconstruction, where the model is trained and tested within seen domains, to establish baseline accuracy in familiar settings. The external assessments include i) cross-center generalization, where the model is evaluated on previously unseen centers to capture institutional heterogeneity; and ii) cross-field-strength generalization, where the model is tested on low-field (0.55T) and ultra-high-field (5.0T) CMR that were absent during training (high-field 1.5T and 3.0T), examining adaptability to different magnetic field strengths.

Furthermore, we design a clinical applicability assessment to examine the value of accelerated CMR image reconstruction in clinical analysis and diagnostic workflows. It includes: i) automated imaging phenotyping, in which accelerated reconstructions are compared with fully sampled references and their diagnostic support is assessed in representative CVDs; and ii) quantitative myocardial biomarkers, where the consistency of key quantitative indices across reconstruction settings is evaluated against fully sampled references and their impact on diagnosis is analyzed. In addition to these objective evaluations, a reader study is performed with experienced radiologists to provide visual scores, offering a complementary clinical perspective on reconstruction reliability.

## Universal reconstruction across internal scenarios

To evaluate the performance of our CardioMM, we conducted extensive internal assessments across eight internal centers using three undersampling patterns (uniform, random, radial) with varying AFs (8×–24×). This assessment involved 75,753 multi-coil k-space data from 1,495 scans of 320 participants, covering 12 CMR modalities acquired on routine high-field scanners (1.5T and 3.0T). For comparison, we included four representative reconstruction methods: a conventional iterative method SENSE[10], widely adopted in commercial scanners, referred to as Conventional in this work; a baseline model DCUNet, which extends a standard UNet[42] with data consistency and coil sensitivity estimation modules[44]; a state-of-the-art universal model PromptMR[39,43], which adapts to diverse scenarios through implicit prompts; and our text-unaware variant CardioSM, designed to directly assess the contribution of our text-aware components in CardioMM. Except for the conventional method, all models were trained on the training subset of MMCMR-427K.

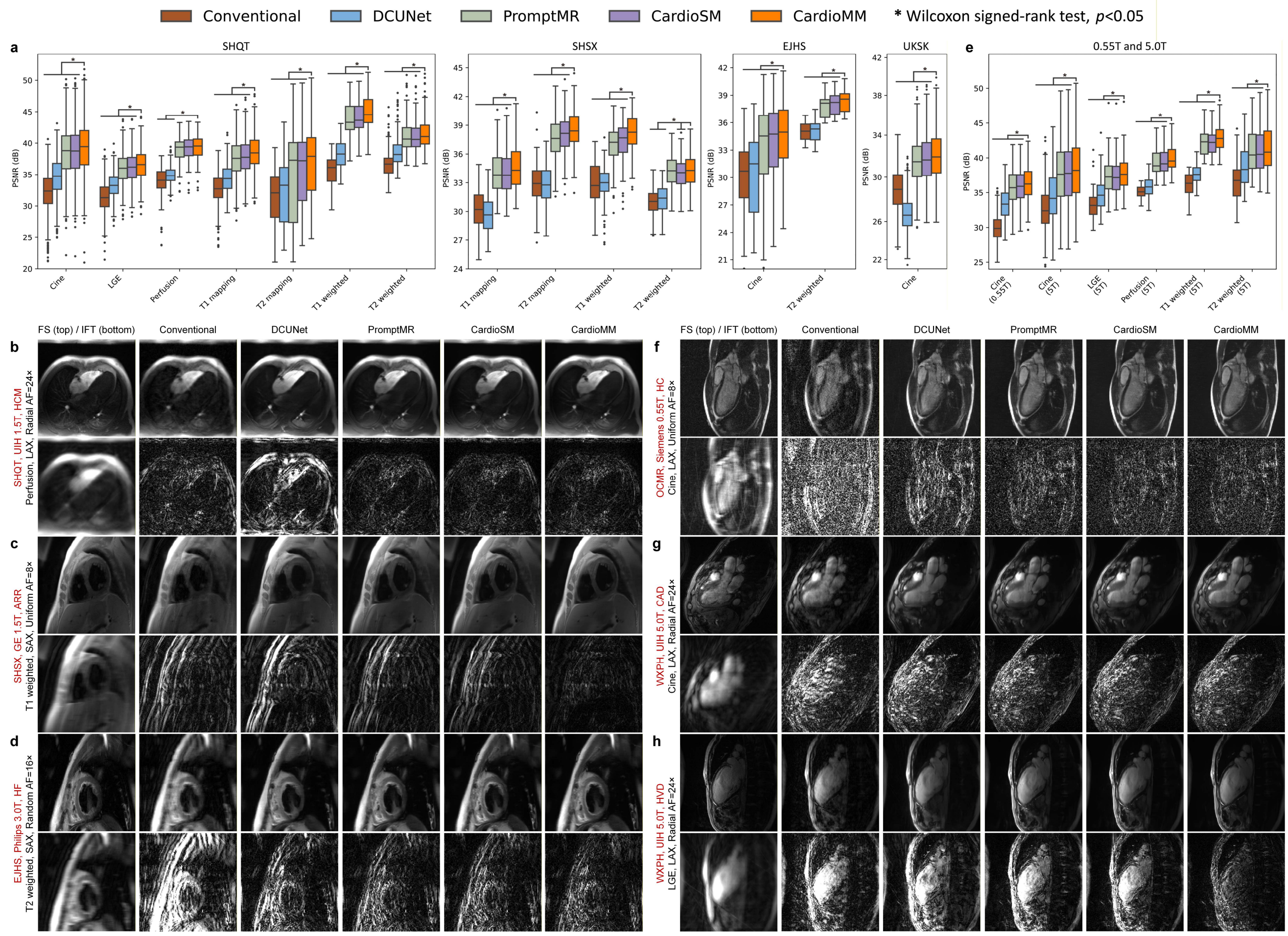

**Fig. 4 | Generalization capability across external centers and field strengths. a**, quantitative comparisons of reconstructions are shown for each modality from each external center, using PSNR. **b-d**, representative reconstruction examples of different methods and their corresponding error maps (scale 0–0.1) from external centers. **e**, quantitative comparisons of reconstructions are shown for each modality from external field strengths, using PSNR. **f-h**, representative reconstruction examples of different methods and their corresponding error maps (scale 0–0.1) from external field strengths. Note: This evaluation is conducted using three undersampling patterns (uniform, random, radial) with varying AFs (4×–24×). The reported median values in the box charts are computed over all tested data for each modality, respectively. "FS" is the fully sampled reference. "IFT" indicates that using only inverse Fourier transform to reconstruct undersampled k-space leads to images with strong artifacts.

We adopted PSNR and SSIM as evaluation metrics here. As shown in Fig. 3 and Supplementary Note 3, our CardioMM consistently outperforms all other compared methods both quantitatively and visually. Large-scale universal models (i.e., CardioMM, CardioSM, and PromptMR) clearly surpass the conventional and baseline methods. Within the universal family, CardioMM achieves the best overall performance with PSNR of 37.94 dB (95% CI: 37.86–38.03 dB) and SSIM of 0.9483 (95% CI: 0.9476–0.9490), averaged over all modalities. This significantly outperforms other text-unaware universal models, with PromptMR obtaining PSNR of 37.15 dB (95% CI: 37.06–37.24 dB) and SSIM of 0.9403 (95% CI: 0.9394–0.9412), and CardioSM obtaining PSNR of 37.26 dB (95% CI: 37.17–37.34 dB) and SSIM of 0.9427 (95% CI: 0.9419–0.9435).

A detailed modality-wise analysis further confirmed the superiority of our CardioMM. Figs. 3a-b show that it outperforms all compared methods across 12 modalities, including the most clinically relevant ones such as cine, LGE, and T1 mapping, with PSNR of 38.82 dB (95% CI: 38.69–38.96 dB), 36.10 dB (95% CI: 35.92–36.28 dB), and 37.06 dB (95% CI: 36.91–37.20 dB), respectively. Consistent gains are also observed in SSIM. Our CardioMM consistently achieved a notable margin over all text-unaware universal models, including the variant CardioSM, while CardioSM fails to suppress PromptMR in some modalities (e.g., T2 weighted, black blood, tagging). It highlights the substantial contribution of the text-aware components in enhancing the multimodal universal reconstruction of our framework.

Representative reconstruction examples are shown in Figs. 3c-h. CardioMM demonstrates strong artifacts suppression, accurate contrast recovery, and faithful preservation of fine structural details, whereas other methods often suffer from residual aliasing, contrast distortion, or loss of cardiac structural information under high accelerations.

These results demonstrate the versatility of CardioMM across diverse centers, modalities, and undersampling scenarios, establishing its strong potential as a universal solution for high-quality multimodal CMR reconstruction under a wide range of ultra-fast imaging requirements.

**Generalization capability across external centers**

Data from different imaging centers often exhibit substantial heterogeneity, largely due to variations in acquisitions, including differences in scanners, imaging protocols, and scan populations[26,45]. Such distribution shifts are particularly common in real-world cardiovascular imaging and impose higher demands on model generalizability[6].

To evaluate this capability, we assessed our CardioMM and other four compared methods on external centers that were not included in training. Specifically, we conducted cross-center evaluations across four external centers using three undersampling patterns (uniform, random, radial) with varying AFs (4×–24×). This evaluation involved 101,069 multi-coil k-space datasets from 1,115 scans of 321 participants, covering seven major CMR modalities acquired on routine high-field scanners (1.5T and 3.0T). These data represented distributions markedly different from those of the internal training centers. Taking the cine modality as example, the training data primarily involved Asian and North American centers, whereas the external evaluation additionally included the UKSK center from Europe[32], introducing clear shifts in scanning and demographic characteristics.

In these external center evaluations, all models were directly tested in a zero-shot setting without any further re-training or fine-tuning, to reflect practical deployment scenarios. Figs. 4a-d and Supplementary Note 4 show that our CardioMM consistently achieves the best zero-shot performance across all external centers and modalities, both quantitatively and visually. For instance, on the European UKSK center, CardioMM reaches PSNR of 32.28 dB (95% CI: 32.15–32.42 dB), significantly surpassing the state-of-the-art PromptMR by 0.57 dB. In contrast, the baseline DCUNet even underperforms the conventional method, with a PSNR drop of up to 9.0%, highlighting the limitations of small-scale models in cross-center generalization and underscoring the necessity of developing large-scale foundation models.

These results demonstrate that CardioMM achieves remarkable zero-shot generalization to unseen centers, scanners, imaging protocols, and study populations, without the need for costly re-training or fine-tuning, thereby highlighting its efficient potential for clinical deployment.

**Generalization capability across external field strengths**

In recent years, CMR has expanded to an unprecedented range of magnetic field strengths[6]. In addition to routine high-field systems, emerging low-field scanners offer advantages such as lower cost and improved patient accessibility[46], while ultra-high-field systems enable higher signal-to-noise ratio (SNR) and novel tissue contrasts[47]. However, these systems inherently differ in SNR and contrast mechanisms, making cross-field-strength generalization a challenging task.

Beyond external center evaluations, we further assessed the performance of our CardioMM under external field strength scenarios. Specifically, we examined its ability to reconstruct CMR data from two previously unseen field strengths (i.e., low-field 0.55T and ultra-high-field 5.0T) across three centers using three undersampling patterns (uniform, random, radial) with varying AFs (8×–24×). It involved 9,117 multi-coil k-space datasets from 110 scans of 74 participants, covering five major CMR modalities.

Figs. 4e-h and Supplementary Note 5 demonstrate that our CardioMM consistently achieves the best zero-shot performance across all modalities at both field strengths and surpasses other methods, both quantitatively and visually. For the 0.55T system, CardioMM reaches the average PSNR of 36.40 dB (95% CI: 35.93–36.86 dB) and SSIM of 0.9070 (95% CI: 0.8987–0.9155). For the 5.0T system, it provides the average PSNR of 38.91 dB (95% CI: 38.60–39.23 dB) and SSIM of 0.9512 (95% CI: 0.9483–0.9543). Notably, under ultra-high acceleration at 5.0T, when all compared methods exhibit severe contrast distortions, our CardioMM still preserves faithful contrast in the cardiac region (Fig. 4h).

These findings demonstrate that CardioMM has strong zero-shot generalization capability across different field strengths, effectively adapting to variations in SNR and contrast. This highlights its broad applicability across emerging low-field, routine high-field, and advanced ultra-high-field CMR systems.

**Clinical applicability of automated imaging phenotyping for diagnostic support**

CMR is the standard imaging tool for the assessment of CVDs. It enables accurate quantification of cardiac structural and functional phenotypes such as ventricular volumes, ejection fraction, and wall thickness (Fig. 5a), thereby providing essential support for the diagnosis and monitoring of multiple CVDs[48].

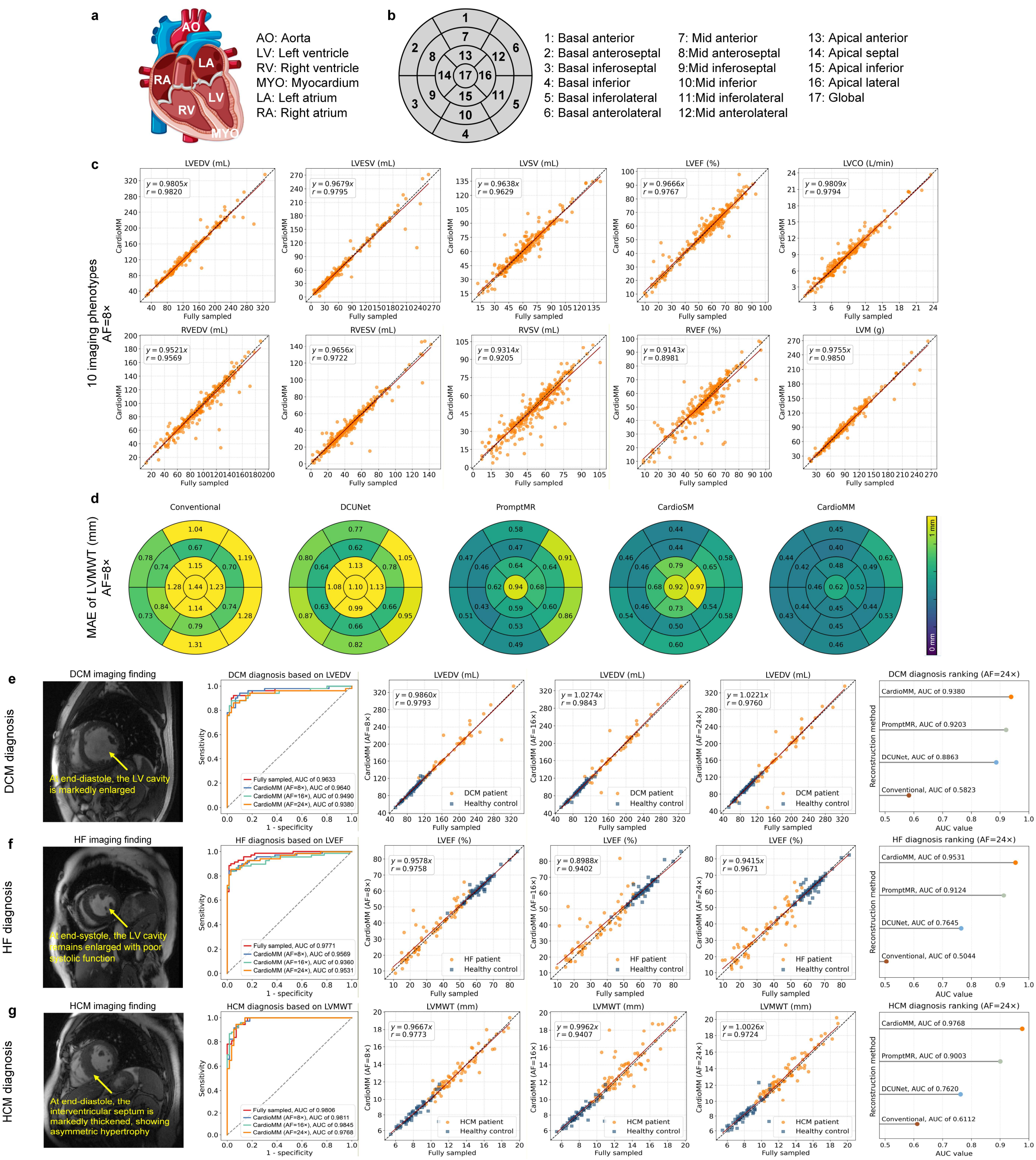

**Fig. 5 | Clinical applicability of automated imaging phenotyping for diagnostic support. a**, schematic illustration of cardiac anatomy. **b**, Bullseye chart of the AHA 16-segment model with a global segment. **c**, linear regression and PCC analysis of 10 representative cardiac imaging phenotypes derived from fully sampled and CardioMM-reconstructed images. **d**, bullseye charts show the average MAE of LVMWT between fully sampled reference and different methods. The above two assessments involve 355 participants with multi-slice short-axis cine modality. **e-g**, diagnostic performance of three cardiac phenotypes derived from fully sampled and CardioMM-reconstructed images under different accelerations. Imaging finding, linear regression, PCC analysis, and method ranking are further given for better visualization. This assessment involves 122 participants (52 DCM patients and 70 HCs) for DCM diagnosis; 149 participants (79 HF patients and 70 HCs) for HF diagnosis; 150 participants (80 HCM patients and 70 HCs) for HCM diagnosis. Note: *r* corresponds to the PCC. LVMWT = left ventricular maximum wall thickness, DCM = dilated cardiomyopathy, HF = heart failure, HCM = hypertrophic cardiomyopathy, HC = healthy control. Some vector images are modified from freepik.com.

Beyond the image quality evaluations described above, we further investigated the clinical applicability of our CardioMM by assessing the consistency of key imaging phenotypes derived from high-acceleration reconstructions compared with their fully sampled references. Additionally, we examined three clinically important CVD conditions, i.e., dilated cardiomyopathy (DCM), heart failure (HF), and hypertrophic cardiomyopathy (HCM), to evaluate whether accelerated reconstructions can preserve the diagnostic utility of CMR phenotyping.

To enable large-scale and efficient CMR analysis, we further integrated CardioMM with a widely recognized automated imaging phenotyping pipeline[48]. This assessment involved 355 participants (including healthy controls and patients with various CVDs) with multi-slice short-axis cine modality across all centers. Fully sampled references were derived by applying the same pipeline to the fully sampled images, ensuring a consistent and unbiased comparison.

First, we evaluated the agreement between CardioMM and fully sampled references across 10 representative imaging phenotypes using linear regression, Pearson correlation coefficient (PCC), and Bland-Altman analysis. Fig. 5c, and Supplementary Figs. 3-4 show that our CardioMM maintains high consistency with references under different accelerations (8×–24×), faithfully reflecting cardiac structure and function. For example, in the case of left ventricular ejection fraction (LVEF), CardioMM achieves PCC of 0.9767 and mean difference of 0.58% (95% LoA: -6.46% to 7.62%) at 8× acceleration. By contrast, conventional method fails to provide clinically meaningful results under the same setting, i.e., PCC of 0.6018 and mean difference of 16.15% (95% LoA: -9.93% to 42.24%). Detailed comparisons are provided in Supplementary Table 6-7, where CardioMM achieves the best overall performance.

Next, we evaluated the mean absolute error of left ventricular maximum wall thickness (LVMWT) between CardioMM and fully sampled references using the American Heart Association (AHA) 16-segment model with a global segment[49], visualized with bullseye charts (Fig. 5b). Fig. 5d and Supplementary Fig. 5 show that, across different AFs (8×–24×), CardioMM consistently achieves small deviations in segmental LVMWT compared with references, with errors less than 1 mm across all segments. It implies superior recovery of myocardial structural details compared with other methods. However, other compared methods already exhibit errors exceeding or approaching 1 mm at 8× acceleration, a deviation that could potentially increase the risk of misdiagnosis in myocardial diseases[50].

Furthermore, we explored the phenotype-based diagnostic support capability of CardioMM compared with fully sampled references across three representative CVDs (i.e., DCM, HF, and HCM), using AUC as the evaluation metric. Among the phenotypes, LVEDV, LVEF, and LVMWT have been shown to provide significant diagnostic value in distinguishing these patient groups from healthy controls, respectively[51,52]. As shown in Fig. 5e and Supplementary Table 8, for LVEDV-based DCM diagnosis, CardioMM maintains diagnostic performance comparable to the references across 8×–24× accelerations. Even in our worst case, CardioMM achieves PCC of 0.9760 and AUC of 0.9380, while the reference AUC of 0.9633. Similarly, for LVEF-based HF diagnosis and LVMWT-based HCM diagnosis, CardioMM consistently obtains high diagnostic accuracy, comparable to the references (Figs. 5f-g and Supplementary Table 8). Detailed results of compared methods can also be found in Supplementary Table 8.

These findings indicate that ultra-fast scans reconstructed by our CardioMM can provide accurate and reliable biventricular imaging phenotypes, substantially reduce acquisition time while preserve high diagnostic and image quality. Remarkably, across three clinically critical CVDs, the phenotypes derived from CardioMM reconstructions exhibit diagnostic performance highly consistent with fully sampled references, underscoring its strong potential as a clinically applicable alternative for ultra-fast CMR imaging.

**Clinical applicability of quantitative myocardial biomarkers for diagnostic support**

Quantitative myocardial biomarkers derived from CMR play a crucial role in characterizing myocardial tissue properties and guiding clinical management of CVDs[4,53,54]. Among them, LGE and T1/T2 mapping are essential for identifying myocardial infarction (MI) and myocarditis (MC). While ultra-fast imaging greatly improves acquisition efficiency, ensuring the quantitative reliability of reconstructed biomarkers is fundamental for clinical translation. Therefore, we further evaluated the consistency between these imaging biomarkers derived from highly accelerated CardioMM reconstructions and those from fully sampled references in disease cohorts, using linear regression, PCC, and Bland-Altman analysis.

First, we assessed MI patients using the LGE modality. Clinically, LGE mass serves as a critical quantitative biomarker for assessing infarct size, viable myocardium, and prognostic risk stratification in MI patients[53]. LGE mass was quantified as the ratio of enhanced myocardium (i.e., MI lesion) to total myocardial mass. Here, the MI lesion was automatically segmented using the well-established full width at half-maximum method, and the full myocardial region was manually annotated. Figs. 6a-d show that our CardioMM maintains high consistency with references under different accelerations (8×–24×), accurately reflecting infarct distribution and LGE mass. Even at 24× acceleration, CardioMM achieves PCC of 0.9441 and mean difference of -0.77% (95% LoA: -4.06% to 2.52%). By contrast, conventional method provides clinically unacceptable results under the same setting, i.e., PCC of 0.7110 and mean difference of 4.94% (95% LoA: -3.11% to 12.99%). Detailed comparisons are provided in Supplementary Figs. 6-7, where CardioMM has the most stable overall performance.

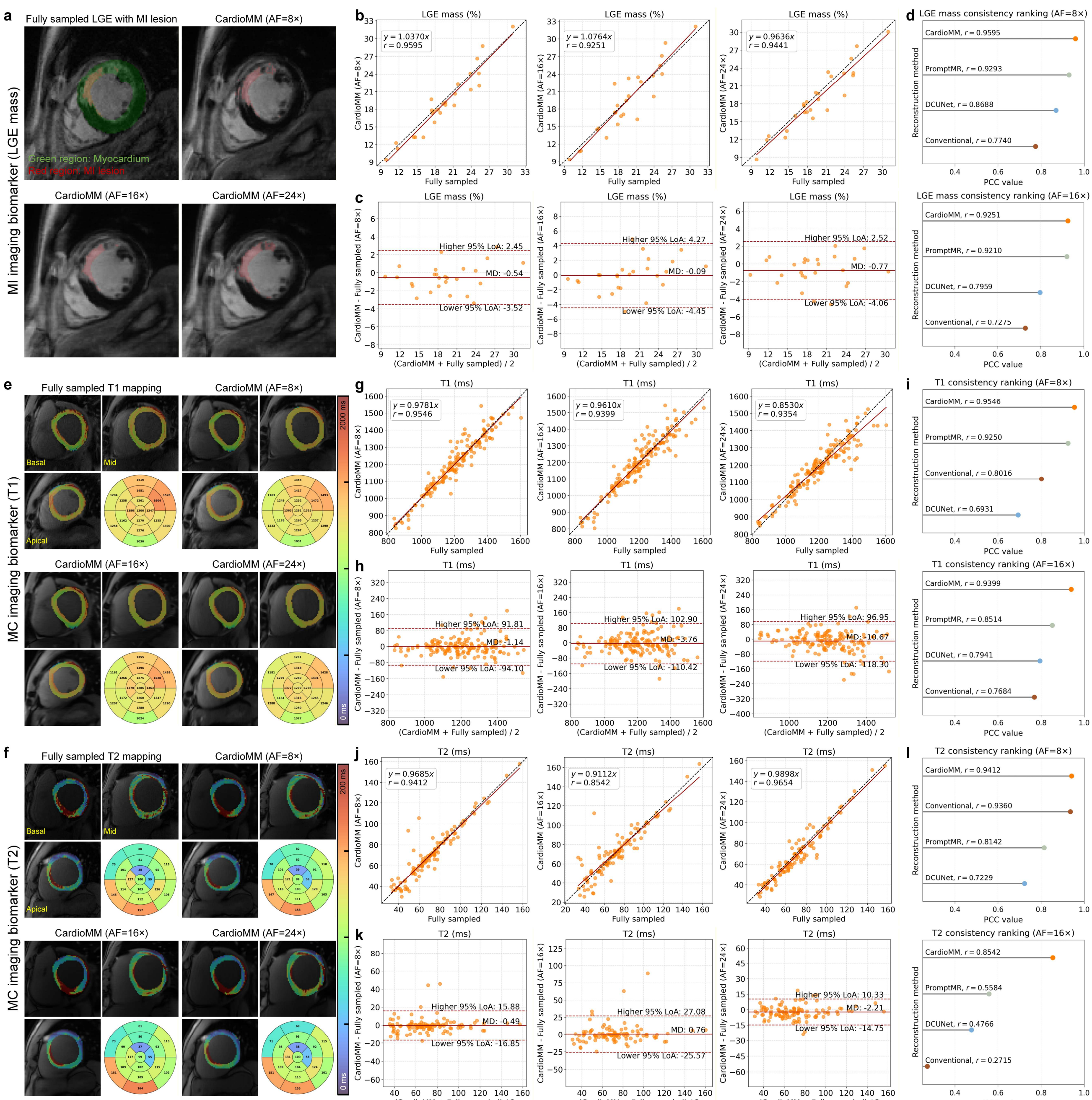

**Fig. 6 | Clinical applicability of quantitative myocardial biomarkers for diagnostic support. a**, representative visualization of MI lesions from fully sampled LGE images and accelerated reconstructions. **b-d**, linear regression, PCC analysis, and Bland-Altman analysis of the MI imaging biomarker (LGE mass) derived from fully sampled and CardioMM-reconstructed images under different accelerations, as well as method ranking. This assessment involves 26 MI patients with multi-slice short-axis LGE modality. **e-f**, representative visualizations and bullseye charts of T1/T2 maps for fully sampled T1/T2 mapping and accelerated reconstructions. **g-l**, linear regression, PCC analysis, and Bland-Altman analysis of the MC imaging biomarker (T1 and T2) derived from fully sampled and CardioMM-reconstructed images under different accelerations, as well as method ranking. This assessment involves 10 MC patients with multi-slice short-axis T1/T2 mapping modalities, and each dot represents a segment-wise T1/T2 value from the AHA 16-segment model. Note: *r* corresponds to the PCC. MD = mean difference. LoA = limits of agreement. MI = myocardial infarction. MC = myocarditis.

Second, for MC patients, we evaluated quantitative T1/T2 values estimated from accelerated CardioMM reconstructions on the T1/T2 mapping. Myocardial T1 and T2 relaxation times are established biomarkers for detecting myocardial inflammation and edema, and elevated T1/T2 values concurrently are critical diagnostic indicators of MC[54]. Here, T1/T2 values were obtained using the least squares fitting method[34], and the myocardial region was manually annotated. Figs. 6e-l show that our CardioMM maintains high consistency with references under different accelerations (8×–24×), accurately providing T1/T2 maps and values. Even at 24× acceleration, CardioMM achieves PCC of 0.9354 for T1 mapping and PCC of 0.9654 for T2 mapping. Additional comparisons with other methods are provided in Supplementary Figs. 8-11. CardioMM consistently delivers the most accurate T1/T2 quantification; whereas some other methods suffer from severe degradation in high-acceleration scenarios, with PCC dropping to as low as 0.6931 for T1 and 0.2715 for T2, leading to MC misdiagnosis.

These results demonstrate that CardioMM enables accurate quantification of key myocardial biomarkers across both structural and parametric modalities, preserving diagnostic reliability under high accelerations. The ability to maintain precise quantitative tissue characterization reinforces the potential of CardioMM for fast and reliable CMR examinations.

### Reader study for qualitative assessment

In clinical practice, accurate diagnosis and interpretation relies not only on the calculation of quantitative CMR metrics but also on expert visual assessment of the images.

Here, we invited five radiologists with 4/4/5/5/6 years' experience, to independently review the reconstructed images from a diagnostic perspective. They were blind to all patient information and reconstruction methods, while fully sampled references were provided. Two clinical-concerned subjective metrics were evaluated: artifacts suppression, and overall image quality. Each metric was rated using a 5-point Likert scale (1: non-diagnostic; 2: poor; 3: adequate; 4: good; 5: excellent). The scores from radiologists were averaged to obtain the final scores of each method. This assessment involved 168 participants with 103 LGE scans, 73 T1 weighted scans, and 88 T2 weighted scans across all available centers.

Supplementary Note 8 shows that our CardioMM scores exceed 4 across all modalities for two metrics. From a diagnostic perspective, its overall image quality was rated between good and excellent (i.e., 4.43 (95% CI: 4.37–4.49)), outperforming other compared methods, making it suitable for clinical diagnosis of multimodal CMR imaging. Notably, even the baseline model DCUNet obtains high scores (i.e., 4.17 (95% CI: 4.11–4.23)) when trained on MMCMR-427K, highlighting that a comprehensive database serves as a critical foundation for multimodal cardiovascular imaging.

### Ablation study

To investigate the effectiveness of the proposed text-aware and dynamic adaptation components, we conducted the ablation study on several model variants with different configurations: i) CardioSM, a text-unaware baseline serving as a purely vision model; ii) CardioSM+UT, which incorporates undersampling texts with the undersampling prompter; and iii) CardioSM+MT, which integrates metadata texts with the metadata adapter.

As summarized in Supplementary Note 9, under internal scenarios, both text-aware variants demonstrate consistent improvements compared with the text-unaware baseline CardioSM. When averaging across all modalities, CardioSM+UT achieves PSNR/SSIM gains of +0.13 dB/+0.08, while CardioSM+MT achieves larger gains of +0.31 dB/+0.23. The superior improvement from metadata-related components suggests that global semantic context plays more important role in guiding multimodal CMR image reconstruction. Most importantly, the full model CardioMM, which jointly employs both the metadata adapter and undersampling prompter, achieves the best overall performance (+0.68 dB/+0.56), clearly surpassing all variants. This highlights the complementary nature of metadata awareness and artifact priors, and demonstrates that their systematic integration is essential for improving both reconstruction accuracy and versatility.

## Discussion and conclusion

High-quality multimodal CMR image reconstruction forms the foundation for all subsequent quantitative and clinical analyses[5,6,15]. This study presents a database–model–validation synergistic paradigm that expands the technological scope of ultra-fast CMR imaging, encompassing the entire pipeline from raw k-space processing to clinically meaningful analysis. By constructing the MMCMR-427K database, the largest and most comprehensive multimodal CMR k-space resource with paired metadata to date, we address one of the most critical bottlenecks in developing generalizable reconstruction models: achieving sufficient data scale, diversity, and semantic completeness. Building upon this infrastructure, we develop CardioMM, a generalist reconstruction foundation model, and demonstrate its capability to achieve high-quality CMR image reconstruction and reliable clinical analysis across heterogeneous imaging environments. This synergistic paradigm further offers a generalizable blueprint for advancing reconstruction foundation models across a wide range of computational imaging fields.

Faster is more than speed—it represents improved imaging robustness. In clinical workflows, multimodal CMR imaging with different structural and functional imaging sequences are routinely acquired to provide complementary diagnostic information. However, this richness comes at the cost of prolonged scan duration, which typically ranges from 30–60 minutes (or even longer), depending on protocol complexity and patient compliance[6]. In time-constrained clinical settings, unavoidable trade-offs must be made among scan efficiency, diagnostic coverage, and image quality. By enabling reliable high-acceleration reconstruction at AFs of 8×–24×, our CardioMM alleviates these limitations and may reshape current clinical scanning paradigms. From large-scale real-world data across heterogeneous centers, we consistently observed that long fully sampled acquisitions often break down in patients with severe CVDs due to breath-hold difficulty and unstable ECG gating, resulting in motion-corrupted k-space measurements, degraded image quality, and compromised diagnostic reliability[5,6,15,26]. Ultra-fast acquisition enabled by CardioMM substantially mitigates this failure mode by shortening the vulnerable time window during which motion and physiological instability corrupt the measurements (Supplementary Fig. 13).

This shifts the role of acceleration from a purely computational improvement in reconstruction metrics to a system-level improvement in acquisition stability and clinical reliability.

Beyond improving workflow efficiency, the ultra-fast multimodal CMR imaging enabled by our CardioMM can expand the applicability of advanced imaging protocols. By shortening the acquisition time of each CMR sequence, additional or more complex sequences, such as mapping and tagging, can be incorporated. This capability enables more comprehensive cardiac characterization within clinically acceptable time windows, facilitating earlier disease detection, more precise lesion delineation, and more personalized treatment planning[5,6]. Moreover, our approach allows the acquisition of richer datasets without extending total scan duration, supporting large-scale cohort studies and longitudinal monitoring, where consistent and fast imaging is essential for tracking disease progression and therapeutic response[32,55,56]. In this way, the synergy between accelerated reconstruction and data-intensive analysis may help bridge the gap between the advanced research and routine clinical practice, advancing the translation toward precision cardiovascular medicine.

Remarkably, previous CMR foundation models mainly focus on post-reconstruction analysis, often assuming the availability of high-quality images from some CMR modalities (e.g., cine and LGE)[9,27,28]. Rather than competing with existing analytical frameworks, our approach complements them by providing higher-quality and more diverse image reconstructions that serve as a robust foundation for downstream segmentation, classification, and phenotyping tasks. Extensive results demonstrate that by integrating text awareness with physics-informed data consistency, our CardioMM achieves a unified balance between semantic authenticity and physical fidelity. Across diverse and previously unseen environments, the model exhibits superior artifact suppression, structural preservation, and zero-shot generalization performance, underscoring its strong potential to handle real-world distribution shifts. Additionally, CardioMM ensures consistent visual, analytical, and diagnostic reliability under varying high AFs (8×–24×), which is a fundamental prerequisite for clinical translation.

The integration of our MMCMR-427K database and our CardioMM model carries significance beyond methodology. With its unprecedented scale and diversity, the database provides a valuable benchmark for studying real-world variability of CMR across institutions and populations. Its paired metadata enables multimodal semantic learning and paves the way for text-conditioned foundation models that integrate imaging physics and contextual knowledge. Such large-scale and standardized resources are crucial to ensuring that AI models encompass diverse demographic and physiological characteristics, which is a key prerequisite for achieving equitable AI applications in healthcare[26].

Despite these advances, several limitations of this study should be acknowledged: i) Our analyses were conducted retrospectively, and prospective deployment within real-time clinical workflows is required to further assess the reliability, speed, and user integration. ii) Although the model demonstrated strong zero-shot generalization to unseen scenarios, further validation is needed for rare disease cohorts, pediatric groups, and patients with implanted devices. iii) The completeness of metadata varies across institutions, and while the frozen text encoder ensures semantic stability, it may limit adaptability to domain-specific terminology. iv) In addition, although the physics-informed framework mitigates hallucination risks, future studies should explore uncertainty quantification, bias assessment, and regulatory compliance to further enhance clinical trustworthiness and ensure diagnostic safety[57].

In the coming era, the synergy between advanced AI and data-driven analysis is likely to become a central axis of precision cardiology. Future work should aim to: i) Expand the MMCMR-427K database by incorporating data from more international collaborators and exploring federated learning and privacy-preserving collaboration frameworks to broaden population diversity without direct data sharing[58]. ii) Develop data-efficient learning strategies, such as self-supervised learning[59], signal-separable learning[41,60,61], and data synthesis[21,22,62], to reduce dependence on paired reference data. iii) Conduct prospective multi-center clinical trials, which are essential for quantifying clinical and economic benefits (e.g., improved throughput and diagnostic reproducibility) and establishing clinician confidence in AI-driven CMR applications.

In conclusion, to the best of our knowledge, this work establishes the first generalist reconstruction foundation model, CardioMM, for ultra-fast multimodal CMR imaging, built on the comprehensive and semantically enriched MMCMR-427K database. It establishes an infrastructure for scalable, generalizable, and high-throughput multimodal cardiovascular imaging. The ability to achieve fast, semantic-aware, and physics-informed image reconstruction not only enhances image quality and diagnostic confidence, but also enables richer data acquisition and large-scale cohort analysis within practical examination time windows.

We anticipate that CardioMM will become a foundational component of next-generation CMR workflows, enabling fast, consistent, and clinically accessible image reconstruction across modalities and centers. More broadly, this study illustrates how foundation modeling can be systematically extended beyond perceptual domains to structured, physics-constrained inverse problems in medical imaging systems, under real-world operator variability and distributional heterogeneity.

## Methods

### Database preparation

Large-scale, diverse, and high-quality databases play a key role in the development of foundation models. In this study, we collected multimodal CMR k-space data from 13 worldwide centers, including four public repositories (OCMR[31], CMRR23[33], CMRR24[34], and UKSK[32]) and nine clinical centers. All real-world clinical data were collected in compliance with ethical standards. The retrospective CMR analysis approved by the institutional review boards, with a waiver of informed consent since no patients were directly recruited or involved. Detailed information of all centers is summarized in Supplementary Table 1.

However, simply aggregating multi-center data is far from sufficient. In clinical practice, CMR acquisition protocols vary widely across centers, resulting in substantial heterogeneity in storage formats and acquisition parameters, which in turn hinders the development of foundation models. To ensure consistency and compatibility of the collected CMR image and text data, we established a unified preprocessing pipeline applied to all centers. This pipeline comprised four major steps: i) k-space standardization, ii) metadata standardization and pairing, iii) demographic characteristics organization and disease classification, and iv) data quality control.

First, in terms of k-space standardization: for the clinical centers, fully sampled k-space references were anonymized by conversion into a raw data format, with all identifiers (e.g., participant name, center location, examination date, and date of birth) removed. The individual k-space lines were sorted according to their acquisition trajectory. To reduce storage demands and computational complexity, coil compression was applied to retain 10 coils for all k-space[63]. The processed k-space was then stored in a unified “mat” format, ensuring consistent dimensional arrangement and facilitating large-scale loading and processing. For the public repositories, a consistent preprocessing and storage procedure was also applied. In particular, since the UKSK center only provided magnitude images without any raw k-space, we synthesized corresponding multi-coil k-space using a physics-informed data synthesis strategy based on the magnitude images[22] (including synthetic phase, coil sensitivities, and Gaussian noise). To establish different acceleration scenarios and reconstruction tasks, various retrospective undersampling patterns (i.e., uniform, random, radial) with AFs ranging from 4× to 24× were generated[34,41]. Undersampling was implemented by retrospectively applying binary masks to fully sampled k-space references. The AF was defined as the ratio of the number of fully sampled k-space data points to the number of acquired points, excluding additional central autocalibration signals (i.e., 20 lines or a 20×20 region).

Second, for metadata standardization and pairing: for the clinical centers, we extracted essential metadata from the corresponding DICOM headers and paired them with the k-space. These metadata included information on acquisition hardware (e.g., vendor, scanner, and field strength) and sequence parameters (e.g., modality, view, resolution, echo time, and repetition time). The processed metadata were then stored in a unified “csv” format, with standardized dimensional arrangement. For the public repositories, we followed the same procedure by utilizing their available metadata and reorganizing them into the standard format.

Third, for demographic characteristics organization and disease classification: for all centers, we collected available demographic information for each participant, including age, sex, height, and weight. CVD information was obtained from the corresponding center episode statistics or clinical records, and classified into 17 categories according to ICD-10 codes[64] (Supplementary Table 2). Participants without any reported CVD were identified as healthy controls.

Finally, data quality control was performed to exclude ineligible data. This step was mainly applied to our clinical centers, as the public repositories had already undergone quality control before release. Quality control was carefully carried out by five radiologists (with 4/4/5/5/6 years’ experience) through systematic visual assessment, and low-quality data with obvious motion, magnetic susceptibility, metal-induced, or aliasing artifacts were excluded.

The resulting MMCMR-427K database was divided into eight internal centers and five external centers (Fig. 1). A total of 241,526 k-space from 3,400 scans of 789 participants were randomly selected from the internal centers for model training, with a 9:1 split between training and validation subsets. The remaining internal center data and all external center data were used to form two test subsets: i) the internal test subset has 75,753 k-space from 1,495 scans of 320 participants, and ii) the external test subset has 110,186 k-space datasets from 1,225 scans of 395 participants. They were used to comprehensively evaluate the model’s performance across diverse test scenarios.

**Implementation of the CardioMM framework**

The proposed CardioMM framework unrolls the iterative reconstruction pipeline into alternating text-aware image de-aliasing modules and physics-informed data consistency modules, enabling high-quality and reliable multimodal CMR image reconstruction guided simultaneously by clinical semantic contexts and underlying imaging physics. The total number of our network phase is empirically set to 10, providing a trade-off between the reconstruction performance and time consumption. The total number of our network parameters is 132M, of which 63M is from a frozen CLIP text encoder (ViT-B/16)[40] for text representation, and the remaining parameters are trainable. Detailed model architecture specifications are provided in Supplementary Note 2, and other hyperparameter settings can be found in our shared codebase.

For model training, we minimized the SSIM loss between fully sampled references and reconstructed images. To enhance robustness, we further developed an automated undersampling generator that dynamically produces diverse undersampling pattern and AF combinations during training, thereby exposing the model to mixed undersampling scenarios. The CardioMM model was trained using the AdamW optimizer with a weight decay of 0.01 for 15 epochs. The initial learning rate was set to 0.0002 and decayed by a factor of 0.3 every five epochs. A batch size of 1 was adopted, to preserve the original spatial dimensions of each k-space without additional cropping, ensuring flexibility in handling varying input sizes and better reflecting the complexity of real-world clinical settings.

The CardioMM framework was implemented in PyTorch 2.0 and trained in parallel across four NVIDIA RTX A6000 GPUs (48 GB memory each) on a server equipped with dual Intel Xeon Gold 6330 CPUs and 502 GB RAM. Typical training on the training subset of our MMCMR-427K database required approximately 7 days. Once trained, the model achieved ultra-fast and generalizable multimodal CMR image reconstruction, with a typical reconstruction time of 0.2 seconds for a multi-coil k-space of size 512×246.

Beyond high-quality multimodal CMR image reconstruction, our CardioMM framework was further integrated with a widely recognized automated imaging phenotyping pipeline[48] to enable large-scale and efficient CMR analysis. This integration supports accurate quantification of 27 representative cardiac structural and functional phenotypes, including ventricular volumes, ejection fraction, and wall thickness, which are widely used for CVD diagnosis and monitoring. The automated phenotyping pipeline consisted of three main steps: i) segmentation of short-axis cine images using a dedicated nnUNet[56,65], automatically delineating the left ventricle (LV), right ventricle (RV), and myocardium (MYO) region (Fig. 4a); ii) automated identification of the end-diastolic (ED) and end-systolic (ES) frames; iii) calculation of 27 phenotypes, including 10 biventricular functional and structural indices (LVEDV, LVESV, LVSV, LVCO, LVM, LVEF, RVEDV, RVESV, RVSV, RVEF), as well as 17 regional LVMWT indices derived from the AHA 16-segment model with an additional global segment[49].

**Evaluation criteria and statistical analysis**

To quantitatively evaluate the reconstruction performance, we employed a combination of objective and subjective evaluation metrics.

For objective reconstruction performance, peak signal-to-noise ratio (PSNR) and the structural similarity index (SSIM)[66] were computed, where higher values indicate fewer image distortions and better structural fidelity, respectively.

For clinical applicability, we assessed the consistency of accelerated reconstructions with fully sampled references using Pearson correlation coefficient (PCC) *r*, mean absolute error (MAE), the area under the receiver operating characteristic curve (AUC)[67], and mean difference (MD) of the Bland-Altman analysis. These metrics reflect the agreement of imaging phenotypes and quantitative myocardial biomarkers with their fully sampled references across different reconstruction settings.

For the reader study, two clinical-concerned subjective metrics including artifacts suppression, and overall image quality were independently rated by experienced radiologists. Each metric was rated using a 5-point Likert scale (1: non-diagnostic; 2: poor; 3: adequate; 4: good; 5: excellent).

For statistical analysis, when the performance differences were tested using the paired two-sided t-test, with $p<0.05$ considered statistically significant. For non-Gaussian data distributions, the Wilcoxon signed-rank test was applied, with $p<0.05$ regarded as statistically significant. The Bootstrap resampling test was also used when appropriate, with $p<0.05$ regarded as statistically significant.

**Compared methods**

We compared the proposed CardioMM with four reconstruction methods: a conventional iterative method SENSE[10], referred to as Conventional in this work; a baseline model DCUNet, which is based on a standard UNet[42] with some modifications for multi-coil k-space processing; a state-of-the-art universal model PromptMR[39,43], that adapts to diverse scenarios through implicit prompts; and our text-unaware variant CardioSM, which is a purely vision model without any text-aware components.

Except for the conventional method, all models were trained on the training subset of our MMCMR-427K database with mixed undersampling scenarios, and then evaluated on different internal and external scenarios without further re-training or fine-tuning.

We included an iterative method SENSE as a conventional baseline since it is widely adopted in commercial scanners. However, it typically supports only relatively low AFs (e.g., ≤3×). Here, we aimed to systematically investigate its reliability for multimodal CMR reconstruction and analysis under higher acceleration settings (e.g., ≥8×). Its implementation was based on the SigPy toolbox[68].

We selected DCUNet as a baseline AI model because it is a representative small-scale reconstruction network. To better handle multi-coil k-space data, it extends a 3-level UNet architecture by incorporating data consistency and coil sensitivity estimation modules[44]. The number of convolutional filters follows 64, 128, 256, and 512 across successive levels.

PromptMR is a state-of-the-art large-scale universal CMR image reconstruction model, which won the championship in the CMRxRecon challenge[29] and has since been widely adopted as a backbone for related tasks[30]. It has the unrolled UNet-like architecture with data consistency and coil sensitivity estimation modules[44], augmented with learnable prompts designed to adapt the model to diverse scenarios. Since the prompts are learned in a data-driven manner, their effectiveness is not guaranteed and the correspondence between data and prompts remains unclear. It was implemented according to the shared code with typical settings.

## Code and data availability

The relevant database, codes, and models will be shared at https://github.com/wangziblake/CardioMM_MMCMR-427K.

All used public datasets are available on their websites, including https://github.com/CmrxRecon, https://ocmr.info, and https://www.ukbiobank.ac.uk. For UK Biobank, the imaging data and non-imaging participant characteristics are available to approved researchers via a standard application process at http://www.ukbiobank.ac.uk/register-apply. Besides, all other clinical CMR datasets from our collection are publicly available.

## Acknowledgements

The authors thank Drs. Jure Zbontar, Anuroop Sriram, Bingyu Xin, Ilya Sutskever, Fabian Isensee, and Devran Ugurlu for sharing their codes online. This research has been conducted using the UK Biobank Resource under Application Number 100203. This study was supported in part by the Shanghai Municipal Science and Technology Major Project (no. 2023SHZD2X02A05), National Natural Science Foundation of China (no. 62331021, 62371413, 62122064, 62201155), Shanghai Rising-Star Program (no. 24QA2703300), Scientific Research Fund Project of Pudong Hospital Affiliated to Fudan University (no. YJJC202409), National Key R&D Program of China (no. 2024YFC3405800), Specialty Feature Construction Project of Pudong Health and Family Planning Commission of Shanghai (no. PWZzb2022-29), Swiss National Science Foundation (no. 220785), ERC IMI (no.101005122), H2020 (no. 952172), MRC (no. MC/PC/21013), Royal Society (no. IEC\NSFC\211235), NVIDIA Academic Hardware Grant Program, SABER project supported by Boehringer Ingelheim Ltd, NIHR Imperial Biomedical Research Centre (no. RDA01), Wellcome Leap Dynamic Resilience, UKRI guarantee funding for Horizon Europe MSCA Postdoctoral Fellowships (no. EP/Z002206/1),

UKRI MRC Research Grant, TFS Research Grants (no. MR/U506710/1), UKRI Future Leaders Fellowship (no. MR/V023799/1), Engineering and Physical Sciences Research Council UK Grants (no. EP/X039277/1), Industry-university Cooperation Projects of the Ministry of Education of China (no. 231107173160805), Yantai Basic Research Key Project (no. 2023JCYJ041), Youth Innovation Science and Technology Support Program of Shandong Provincial (no. 2023KJ239), Youth Program of Natural Science Foundation of Shandong Province (no. ZR2024QF001), Shanghai Science and Technology Commission "Explorer Project" (no. 24TS1410400), Imperial College London Seeds for Success Fund, and Imperial College London I-X.

## Author contributions

C. Wang and Z. Wang conceived the idea and designed the study. Z. Wang, M. Huang, and X. Qu developed the method, constructed the database, performed the experiments, and analyzed the data. Z. Wang and M. Huang prepared all figures and tables for the manuscript and supplementary materials. Z. Shi, H. Hu, Lan Lan, and Hui Zhang contributed to clinical data acquisition and curation, and provided critical revisions to the result interpretation. Y. Li, Xi Hu, Q. Lu, Z. Zhu, F. Wang, Y. Wu, Q. Gao, G. Xu, Z. Zhang, Z. Xu, Q. Yao, L. Xue, Y. Lyu, J. Zhu, R. Ahmad, Z. Bu, X. Qian, F. Yu, S. Ma, G. Cai, S. Hua, Y. Zhang, L. Wu, M. Zeng, Xihong Hu, and H. Xu assisted with data acquisition and processing, and provided suggestions on database construction. Y. Dai, Haosen Zhang, Q. Li, G. Wang, T. He, Lizhen Lan, S. Li, M. Sun, J. Hu, R. Cao, W. Cai, C. Xu, X. Chen, J. Qin, Y. Yang, J. Lyu, C. Qin, S. Wang, C. Ouyang, D. Kim, W. Bai, H. Wang, Q. Tao, D. Rueckert, C. Prieto, M. Markl, A. Young, X. Qu, H. Li, G. Yang, and C. Wang provided methodological suggestions and critical revisions to the manuscript. C. Wang, G. Yang, H. Li, X. Qu, H. Xu, and Xihong Hu supervised the project and provided all necessary resources. The manuscript was drafted by Z. Wang, and all authors discussed the results, contributed revisions, and reviewed the final manuscript.

## Competing interests

The authors declare no competing interests.

# Supplementary Material for

# "Enabling Ultra-Fast Cardiovascular Imaging Across Heterogeneous Clinical Environments with A Generalist Foundation Model and Multimodal Database"

## Supplementary Note 1. MMCMR-427K database

**Supplementary Table 1 | Detailed description and characteristics of our MMCMR-427K database, containing 427,465 multi-coil k-space data from 6,120 scans of 1,504 participants across 13 centers.**

| Center | Population | Age / BMI (mean±std) | Disease | Scanner | Participant number | Modality | Scan number | Paired k-space and metadata number |
|---|---|---|---|---|---|---|---|---|
| Internal center | | | | | | | | |
| RJHE | Asian | 47±15 years<br>/<br>23.90±4.33 | HC<br>CAD<br>HCM<br>DCM<br>UCM<br>MC<br>PC<br>HHD<br>ARR<br>HF<br>HVD<br>CHD | 3.0T Siemens Vida<br>3.0T UIH uMR780 | 91<br>/<br>Male 41<br>Female 50 | Cine<br>T1 mapping<br>T2 mapping<br>LGE<br>Perfusion<br>T1rho mapping<br>T2 weighted | 89<br>87<br>69<br>46<br>46<br>58<br>71 | 12,252<br>3,305<br>1,083<br>580<br>9,035<br>803<br>793 |
| ZSHFD | Asian | 58±15 years<br>/<br>24.25±3.94 | CAD<br>HCM<br>MI<br>ARR | 3.0T Siemens Cima.X<br>3.0T UIH uMR880 | 30<br>/<br>Male 18<br>Female 12 | Cine<br>T1 mapping<br>T2 mapping<br>LGE<br>Perfusion<br>T1rho mapping<br>T2 weighted | 29<br>20<br>25<br>27<br>3<br>10<br>2 | 1,812<br>738<br>398<br>564<br>600<br>128<br>18 |
| SHGC | Asian | 55±14 years<br>/<br>23.84±3.24 | HC<br>CAD<br>HCM<br>DCM<br>UCM<br>MC<br>PC<br>HHD<br>PAH<br>ARR<br>HF<br>HVD<br>CHD | 1.5T UIH uMR670<br>3.0T UIH uMR880 | 58<br>/<br>Male 40<br>Female 18 | Cine<br>T1 mapping<br>T2 mapping<br>LGE<br>T1 weighted<br>T2 weighted | 58<br>25<br>40<br>43<br>42<br>4 | 6,660<br>1,195<br>360<br>1,008<br>418<br>40 |
| SHQC | Asian | 54±16 years<br>/<br>23.84±3.87 | HC<br>CAD<br>HCM<br>DCM<br>RCM<br>UCM<br>MC<br>MI<br>PC<br>HHD<br>PAH<br>ARR<br>HF | 1.5T UIH uMR670<br>1.5T GE Voyager<br>3.0T Siemens Vida | 188<br>/<br>Male 104<br>Female 84 | Cine<br>T1 mapping<br>T2 mapping<br>LGE<br>Perfusion<br>T1 weighted<br>T2 weighted | 112<br>185<br>183<br>110<br>67<br>174<br>169 | 13,428<br>8,138<br>1,890<br>2,570<br>13,350<br>1,758<br>1,647 |
| | | | HVD<br>CHD<br>CBN<br>CMN<br>CAM | | | | | |
| ZNHWH | Asian | 41±20 years<br>/<br>27.79±3.65 | HC<br>CAD<br>HCM<br>DCM<br>UCM<br>MC<br>MI<br>PC<br>HHD<br>PAH<br>ARR<br>HVD<br>CHD | 3.0T Siemens Prisma<br>3.0T UIH uMR790<br>5.0T UIH uMRJupiter | 93<br>/<br>Male 47<br>Female 46 | Cine<br>T1 mapping<br>T2 mapping<br>LGE<br>2D flow<br>Black blood | 93<br>46<br>49<br>39<br>36<br>19 | 14,664<br>2,077<br>366<br>686<br>1,454<br>155 |
| OCMR[1] | North American | N/A | HC | 0.55T Siemens Free.Max<br>1.5T Siemens Avanto<br>1.5T Siemens Sola<br>3.0T Siemens Prisma<br>3.0T Siemens Vida | 78<br>/<br>Male N/A<br>Female N/A | Cine | 78 | 2,628 |
| CMRR23[2] | Asian | 26±5 years<br>/<br>N/A | HC | 3.0T Siemens Vida | 300<br>/<br>Male 140<br>Female 160 | Cine<br>T1 mapping<br>T2 mapping | 274<br>287<br>286 | 39,756<br>13,950<br>4,632 |
| CMRR24[3] | Asian | 36±12 years<br>/<br>23.35±3.46 | HC | 3.0T Siemens Vida | 330<br>/<br>Male 174<br>Female 156 | Cine<br>T1 mapping<br>T2 mapping<br>2D flow<br>Black blood<br>Aorta<br>Tagging | 326<br>321<br>322<br>250<br>245<br>249<br>240 | 52,176<br>15,633<br>5,226<br>6,000<br>1,329<br>46,836<br>31,188 |
| External center | | | | | | | | |
| SHQT | Asian | 51±17 years<br>/<br>29.62±3.88 | HC<br>CAD<br>HCM<br>DCM<br>RCM<br>UCM<br>MC<br>MI<br>PC<br>HHD<br>PAH<br>ARR<br>HF<br>HVD<br>CHD<br>CBN<br>CAM | 1.5T Siemens Aera<br>1.5T UIH umr680 | 175<br>/<br>Male 114<br>Female 61 | Cine<br>T1 mapping<br>T2 mapping<br>LGE<br>Perfusion<br>T1 weighted<br>T2 weighted | 173<br>158<br>156<br>135<br>87<br>78<br>87 | 25,212<br>7,825<br>1,605<br>3,012<br>21,606<br>785<br>868 |
| SHSX | Asian | 56±17 years<br>/<br>N/A | HC<br>CAD<br>HCM<br>DCM<br>RCM<br>UCM<br>PC<br>HHD<br>ARR<br>HF<br>HVD<br>CHD<br>CBN | 1.5T GE Voyager | 32<br>/<br>Male 24<br>Female 8 | T1 mapping<br>T2 mapping<br>T1 weighted<br>T2 weighted | 31<br>32<br>31<br>31 | 1,440<br>384<br>324<br>328 |
| WXPH | Asian | 49±19 years<br>/<br>22.99±2.65 | HC<br>CAD<br>HCM<br>PAH<br>ARR<br>HF | 5.0T UIH uMRJupiter | 15<br>/<br>Male 8<br>Female 7 | Cine<br>LGE<br>Perfusion<br>T1 weighted<br>T2 weighted | 15<br>9<br>9<br>8<br>10 | 876<br>160<br>1,850<br>85<br>98 |

| | | | HHD<br>HVD | | | | | |
| --- | --- | --- | --- | --- | --- | --- | --- | --- |
| EJHS | Asian | 55±14 years<br>/<br>N/A | HC<br>HCM<br>DCM<br>RCM<br>PC<br>PAH<br>HF<br>HVD | 3.0T Philips<br>IngeniaCX | 14<br>/<br>Male 7<br>Female 7 | Cine<br>T2 weighted | 14<br>2 | 3,012<br>18 |
| UKSK[4] | European | N/A | N/A | 1.5T Siemens Aera | 100<br>/<br>Male N/A<br>Female N/A | Cine | 100 | 34,650 |

Note: [1]Available at https://ocmr.info. [2]Available at https://github.com/CmrxRecon/CMRxRecon-SciData. [3]Available at https://github.com/CmrxRecon/CMRxRecon2024. [4]Available at https://www.ukbiobank.ac.uk. UKSK denotes the UK Biobank synthetic k-space, which is generated from the magnitude-only images provided by the UK Biobank using a physics-informed data synthesis strategy[5], including the simulation of phase, coil sensitivities, and measurement noise. Others are clinical centers. HC = healthy control. BMI = body mass index. “N/A” represents information not available or not collected. All cardiovascular diseases are given in abbreviations here, while their full names and detailed information are provided in Supplementary Table 2.

**Supplementary Table 2 | Cardiovascular disease (CVD) categories involved in this study, and one participant may have more than one CVD.**

| CVD abbreviation | CVD | ICD-10 code | CVD case number |
|---|---|---|---|
| CAD | Coronary artery disease | I25 | 75 |
| HCM | Hypertrophic cardiomyopathy | I42.1 | 206 |
| DCM | Dilated cardiomyopathy | I42.0 | 118 |
| RCM | Restrictive cardiomyopathy | I42.5 | 3 |
| UCM | Unspecified cardiomyopathy | I42.9 | 41 |
| MC | Myocarditis | I40 | 19 |
| MI | Myocardial infarction | I21–I22 | 35 |
| PC | Pericarditis | I31 | 77 |
| HHD | Hypertensive heart disease | I11 | 46 |
| PAH | Pulmonary arterial hypertension | I27.0–I27.2 | 9 |
| ARR | Arrhythmia | I47–I49 | 81 |
| HF | Heart failure | I50 | 176 |
| HVD | Heart valve disease | I34–I38 | 197 |
| CHD | Congenital heart disease | Q20–Q28 | 16 |
| CBN | Cardiac benign neoplasm | D15.1 | 3 |
| CMN | Cardiac malignant neoplasm | C38.0 | 1 |
| CAM | Cardiac amyloidosis | I43.1 | 7 |

# Supplementary Note 2. CardioMM methodology

In this section, we first introduce the overall network architecture of the proposed CardioMM, which involves the text encoder with projection heads for text representation, and alternating text-aware image de-aliasing modules and physics-informed data consistency modules (Supplementary Fig. 1). This design ensures that multimodal cardiovascular magnetic resonance (CMR) image reconstruction is guided by both clinical semantic contexts and underlying imaging physics, thereby enhancing the reliability and clinical applicability of the reconstructed outcomes.

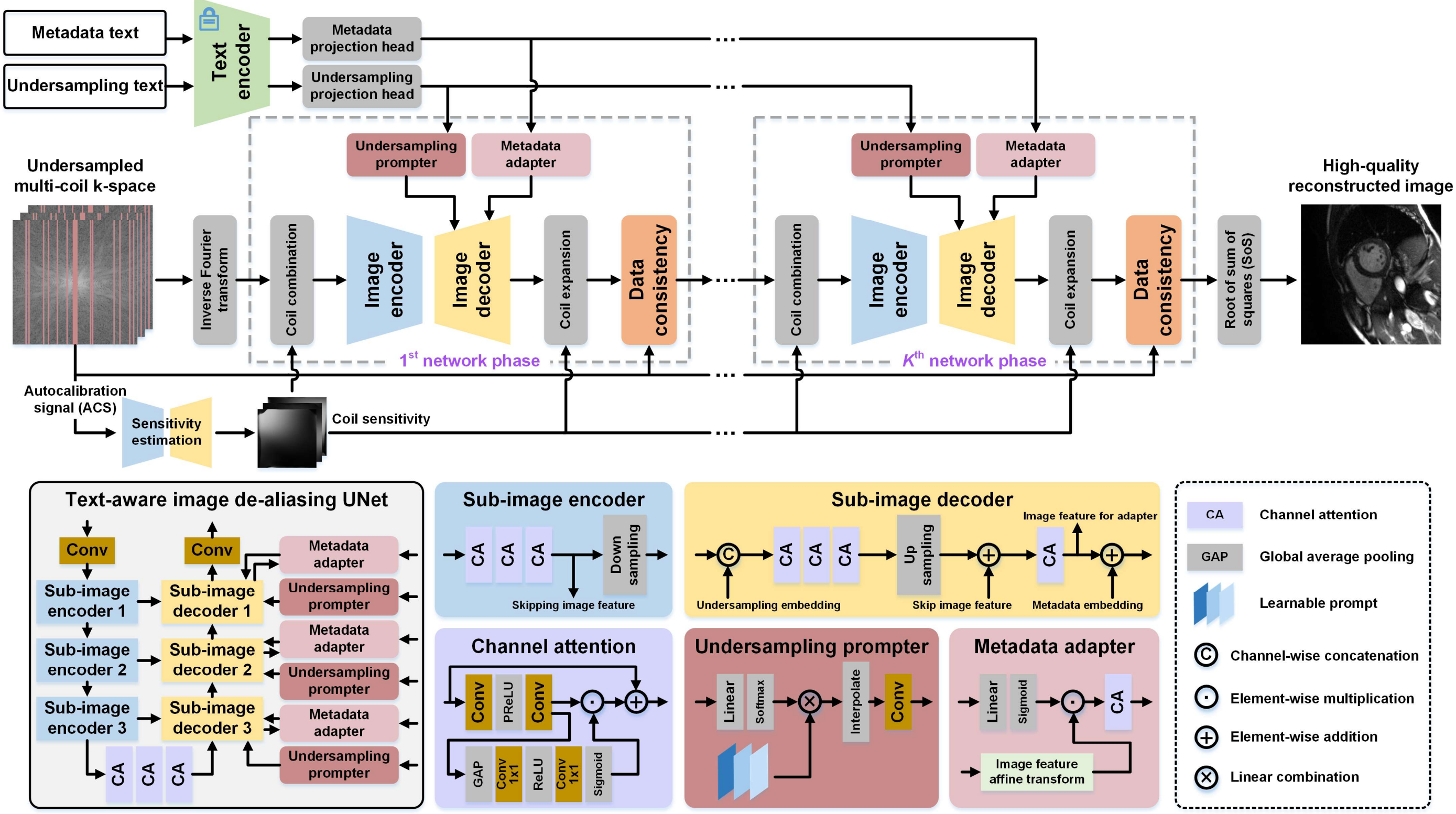

**Supplementary Fig. 1 | The network architecture of the proposed CardioMM for text-aware multimodal cardiovascular image reconstruction.** The detailed structures of the network modules and some definitions are given below the overall pipeline. Note: "ACS" is the fully sampled low-frequency region at the central k-space, which commonly serves as a calibration for coil sensitivity estimation. "SoS" means that the reconstructed multi-coil images are finally displayed after combining by the square root of sum of squares.

## 2.1 Overall Network Architecture

Here, we first formulate the reconstruction model of the vectorized multi-coil image $\mathbf{x}$ with the learned deep image prior:

$$\min_{\mathbf{x}} \left\|\mathbf{y}-\mathcal{U}\mathcal{F}\mathbf{x}\right\|_2^2+\lambda\left\|\mathbf{x}-\mathcal{M}(\mathbf{x},\mathbf{S},\mathbf{t}_M,\mathbf{t}_U)\right\|_2^2, \qquad (1)$$

where $\mathcal{M}$ is the learned text-aware image de-aliasing module, $\lambda$ is the regularization parameter,

$\mathbf{S}=[\mathbf{S}_1;\ldots;\mathbf{S}_j;\ldots\mathbf{S}_J]$ is the set of coil sensitivity maps and $\mathbf{S}_j$ is a diagonal matrix which denotes the sensitivity map of the $j^{\text{th}}$ coil. $\mathbf{t}_M$ and $\mathbf{t}_U$ are text representations from metadata text and undersampling text, respectively. The (1) can be mainly solved by alternating two sub-problems[5,6], and the $k^{\text{th}}$ iteration is:

$$\begin{cases}\mathbf{m}^{(k)}=\mathcal{M}(\mathbf{x}^{(k-1)},\mathbf{S},\mathbf{t}_M,\mathbf{t}_U)\\ \mathbf{x}^{(k)}=\arg\min_{\mathbf{x}}\left\|\mathbf{y}-\mathcal{U}\mathcal{F}\mathbf{x}\right\|_2^2+\lambda\left\|\mathbf{x}-\mathbf{m}^{(k)}\right\|_2^2\\ \quad\;\;=(\mathcal{F}^*\mathcal{U}^*\mathcal{U}\mathcal{F}+\lambda)^{-1}(\mathcal{F}^*\mathcal{U}^*\mathbf{y}+\lambda\mathbf{m}^{(k)})\end{cases}, \tag{2}$$

where $\mathbf{y}$ is the vectorized undersampled multi-coil k-space, the superscript * represents the adjoint operation, $\mathcal{F}$ and $\mathcal{F}^*$ are the Fourier transform and inverse Fourier transform, respectively.

Once the overall number of iterations $K$ is fixed, the iteration process in (2) can be viewed as an unrolled deep network with $K$ phase (Supplementary Fig. 1). Except for the text representation modules, each network phase mainly consists of two modules: A text-aware image de-aliasing module and a physics-informed data consistency module, which correspond to the first and second step of (2), respectively. The final reconstructed multi-coil image is displayed after combining by the square root of sum of squares (SoS). We perform end-to-end training using the large-scale and diverse datasets to learn the model weights and set $\lambda$ as a trainable parameter. If the regularization can yield improved reconstructions, high values of $\lambda$ would be learned during the training process. When $k=1$, the initialized input $\mathbf{x}^{(0)}=\mathcal{F}^*\mathcal{U}^*\mathbf{y}$ is the zero-filled multi-coil image with strong artifacts.

## 2.2 Text representation module

The text encoder transforms original textual information into fixed-size vector representations, known as the text representation. The text encoder from the Contrastive Language-Image Pre-training (CLIP) model[7] is frequently employed to encode textual information, as CLIP demonstrates strong capabilities in capturing underlying semantic information. Although CLIP is mainly trained on natural image-text pairs (some of which may be medically relevant), it can be effectively adapted to specific medical imaging applications (such as classification[8], segmentation[9], and generation[10]), leveraging its zero-shot capabilities either directly or through appropriate fine-tuning[11]. This insight motivates our use of the CLIP text encoder.

Here, we aim to adapt the text encoder $\mathcal{T}$ for the multimodal CMR image reconstruction task to better encode the diverse and complex textual information required by the reconstruction model. Directly training the full text encoder on our specific CMR dataset, which is relatively limited in scale compared to the large corpus used to pretrain CLIP model[7], risks overfitting and loss of generalizability. Therefore, we freeze the parameters of the CLIP text encoder and instead train lightweight projection heads jointly with the reconstruction model, allowing them to learn task-specific text representations.

Specifically, the input text information is divided into two categories: metadata text and undersampling text. The metadata text includes patient and scan-related information such as life stage, imaging protocol, and scanner configuration, which provide critical semantic context for understanding the image itself[2,3]. The undersampling text represents acquisition-specific parameters, such as sampling patterns and acceleration factors (AFs), which relate to the characteristics (i.e., distribution and intensity) of undersampling-induced artifacts[12]. Since image artifacts primarily depend on both the intrinsic image content and the undersampling scenario, encoding these two types of text is decisive for clearly guiding the model to understand and remove image artifacts. Both types of text inputs are processed by the shared frozen text encoder to produce raw text representations. Subsequently, two separate projection heads transform this raw representation into specialized representations tailored for metadata and undersampling scenarios, respectively (Supplementary Fig. 1). This process can be formulated as follows:

$$\mathbf{t}_M = \mathcal{H}_M(\mathcal{T}(\mathbf{m})),\ \mathbf{t}_U = \mathcal{H}_U(\mathcal{T}(\mathbf{u})), \tag{3}$$

where $\mathbf{m}$ and $\mathbf{u}$ are the metadata and undersampling texts, respectively. $\mathcal{H}_M$ and $\mathcal{H}_U$ are metadata and undersampling projection heads, respectively. Each projection head consists of a linear layer followed by L2-normalization. $\mathbf{t}_M$ and $\mathbf{t}_U$ are metadata and undersampling representations, respectively, and are shared across all network phases.

Our design is mainly based on three considerations: 1) Freezing the text encoder reduces the trainable parameters and preserves the board semantic knowledge from large-scale pretraining. 2) Employing distinct projection heads enables task-specific representations that better capture the unique semantics of each text type. 3) Sharing the main text encoder while decoupling the projection heads provides flexibility, facilitating extension to additional text information without re-training the entire module.

## 2.3 Text-aware image de-aliasing module

The text-aware image de-aliasing module is composed of five components: The coil combination operator, text-aware UNet[13], metadata adapter, undersampling prompter, and coil expansion operator. This module takes in a multi-coil undersampled image and aims to recover a high-quality image through adaptive artifact removal that incorporates both semantic and acquisition-specific cues (Supplementary Fig. 1).

To support coil combination and expansion during reconstruction, we first estimate coil sensitivity maps, which are essential for transforming multi-coil images into coil-combined images and vice versa. These coil sensitivity maps are computed by a sensitivity estimation module $\mathcal{S}$ from the autocalibration signal $\mathbf{y}_{ACS}$, which is the fully sampled low-frequency region at the central k-space[14,15]. To be more intuitive, the text-aware image de-aliasing module shown in the first sub-problem of (2) is further decomposed as:

$$\begin{cases} \mathbf{m}^{(k)} = \mathcal{M}(\mathbf{x}^{(k-1)}, \mathbf{S}, \mathbf{t}_M, \mathbf{t}_U) = \mathcal{E}(\mathcal{D}(\mathcal{C}(\mathbf{x}^{(k-1)}, \mathbf{S}^*), \mathbf{t}_M, \mathbf{t}_U), \mathbf{S}) \\ \mathbf{S} = \mathcal{S}(\mathcal{F}^* \mathbf{y}_{ACS}) \end{cases}, \tag{4}$$

where $\mathcal{C}$ is the coil combination operator, $\mathcal{D}$ is the text-aware UNet, and $\mathcal{E}$ is the coil expansion operator. Specifically,

$$\begin{cases} \mathbf{x}_C^{(k)} = \mathcal{C}(\mathbf{x}^{(k-1)}, \mathbf{S}^*) = \sum_{j=1}^{J} \mathbf{S}_j^* \mathbf{x}_j^{(k-1)} \\ \mathbf{x}_D^{(k)} = \mathcal{D}(\mathbf{x}_C^{(k)}, \mathbf{t}_M, \mathbf{t}_U) \\ \mathbf{m}^{(k)} = \mathcal{E}(\mathbf{x}_D^{(k)}, \mathbf{S}) = \mathbf{S}\mathbf{x}_D^{(k)} \end{cases}. \tag{5}$$

All coil sensitivity maps are normalized to satisfy $\sum_{j=1}^{J} \mathbf{S}_j^* \mathbf{S}_j = \mathbf{I}$, where $\mathbf{I}$ is an identity matrix. The sensitivity estimation module $\mathcal{S}$ shares the network architecture to $\mathcal{D}$ but receives different types of input data.

Trained on large-scale and diverse CMR datasets, our network leverages text representations to remove artifacts caused by undersampling. To exploit the complementary nature of two types of textual inputs, we design two separate text-injection mechanisms: 1) Metadata adapter, which introduces global semantic context into the image feature stream in a stable and lightweight manner.

2) Undersampling prompter, which modulates the network's intermediate layers using acquisition-specific information directly related to artifact characteristics. The obtained metadata and undersampling embeddings are injected into the image decoders of our text-aware UNet (Supplementary Fig. 1).

### 2.3.1 Metadata adapter

The metadata adapter is responsible for integrating high-level semantic information, such as patient condition, anatomical region, and imaging configuration, into the image reconstruction process. These attributes modulate image texture, contrast, and structural details, guiding the network's attention toward salient information and influencing the final reconstructed image.

At each UNet level (Supplementary Fig. 1), the metadata representation $\mathbf{t}_M$ is first passed through a linear layer followed by a Sigmoid activation to produce a global modulation weight $\mathbf{w}_M$. The intermediate image feature from the image decoder $\mathbf{f}_A$ is modulated by an affine transformation (i.e., linear modulation)[16,17], followed by scaling with $\mathbf{w}_M$, and further enhanced by a channel attention block[18] $\mathcal{N}_{CA}$ to produce the final metadata embedding $\mathbf{e}_M$. This embedding is then passed into the image decoder pathway of our UNet to guide the image outcomes.

The entire procedure in our metadata adapter can be clearly summarized as:

$$\begin{cases} \mathbf{w}_M^{(k)} = Sigmoid(\mathcal{N}_{Linear}(\mathbf{t}_M)) \\ \mathbf{f}_{AT}^{(k)} = \gamma^{(k)}\mathbf{f}_A^{(k)} + \beta^{(k)} \\ \mathbf{e}_M^{(k)} = \mathcal{N}_{CA}(\mathbf{w}_M^{(k)} \odot \mathbf{f}_{AT}^{(k)}) \end{cases}, \quad (6)$$

where $\gamma$ and $\beta$ are the parameters for the linear modulation, and they are initialized to 1 and 0, respectively. $\odot$ represents the element-wise multiplication.

Such a design achieves two main functions: 1) Global semantic awareness, allowing the network to better understand what and where to look for image features of interest. 2) Adaptive modulation, enabling metadata-aware processing that adjusts to varying imaging scenarios, thereby improving generalizability across patient conditions and imaging protocols. It ensures that our image decoder is dynamically informed by high-level imaging context.

### 2.3.2 Undersampling prompter

The undersampling prompter captures local artifact priors introduced by specific undersampling settings. Since the nature of undersampling (e.g., sampling patterns and AFs) fundamentally shapes the aliasing behavior in the image, we explicitly prompt the network on such information to achieve undersampling-aware reconstruction. To achieve this, the undersampling prompter is introduced at each level of our text-aware UNet and performs the operations in Supplementary Fig. 1.

We first feed the undersampling representation $\mathbf{t}_U$ to a linear layer followed by a Softmax activation to obtain the soft attention weight $\mathbf{w}_U$. Meanwhile, the prompt dictionary $\mathbf{p}_D$ with $Q$ components is maintained[19-21], from which the composite prompt is constructed as a weighted sum $\mathbf{p}_U$. To integrate the prompt into the reconstruction pipeline, we first upsample $\mathbf{p}_U$ using bilinear interpolation to match the spatial resolution of the current image decoder level, then input it into a simple convolutional layer $\mathcal{N}_{Conv}$ to obtain the final undersampling embedding $\mathbf{e}_U$. This embedding is then fused into the image decoder pathway of our UNet to enable prompt injection.

The full process in our undersampling prompter can be formally expressed as:

$$\begin{cases} \mathbf{w}_U^{(k)} = Softmax(\mathcal{N}_{Linear}(\mathbf{t}_U)) \\ \mathbf{p}_U^{(k)} = \mathbf{w}_U^{(k)} \otimes \mathbf{p}_D^{(k)} = \sum_{q=1}^{Q} \mathbf{w}_{U,q}^{(k)} \odot \mathbf{p}_{D,q}^{(k)} \\ \mathbf{e}_U^{(k)} = \mathcal{N}_{Conv}(Interpolate(\mathbf{p}_U^{(k)})) \end{cases}, \tag{7}$$

where $\otimes$ represents the linear combination (i.e., weighted sum) here. In this work, the number of prompt components $Q$ is set to 3, corresponding to three widely used undersampling patterns (i.e., uniform, random, and radial).

This design enables two complementary effects: 1) Artifact-aware prompt, by encoding acquisition-specific priors into prompts that explicitly inform the network how artifacts manifest under varying undersampling scenarios. 2) Multi-level prompt injection, by embedding these prompts at different levels of our image decoder, allowing artifact suppression across spatial resolutions.

### 2.3.3 Text-aware UNet architecture

The backbone of the image de-aliasing module is a 3-level UNet[13] composed of residual connections and channel attention mechanisms[20,21], designed to progressively extract and refine features from undersampled images. To effectively incorporate both semantic context and acquisition-specific prompts, we enhance this vanilla architecture with a dual-text embedding strategy to obtain a new text-aware UNet (See Supplementary Fig. 1): metadata adapters and undersampling prompters are inserted at each level of image decoders. Besides, to preserve generality in the learned image features, text representations are injected only into the decoder. This allows the encoder to focus on capturing a universal representation of the underlying image content, while the decoder dynamically adjusts its outputs according to task-specific textual guidance.

Each image encoder level comprises three channel attention blocks $\mathcal{N}_{CA}$ followed by a downsampling operator. Let $\mathbf{f}_{EI}$ denote the input feature of the encoder. Before downsampling, this skip feature $\mathbf{f}_{S}$ is preserved and passed to the corresponding decoder level via residual connections. This process can be summarized as:

$$\begin{cases} \mathbf{f}_{S}^{(k)} = \mathcal{N}_{CA}(\mathbf{f}_{EI}^{(k)}) \\ \mathbf{f}_{EO}^{(k)} = Downsampling(\mathbf{f}_{S}^{(k)}) \end{cases} . \tag{8}$$

The image decoder incorporates both metadata and undersampling embeddings at each level. Specifically, each decoder level involves: 1) Concatenation of the undersampling embedding $\mathbf{e}_{U}$ and the current decoder input $\mathbf{f}_{DI}$, followed by three channel attention blocks $\mathcal{N}_{CA}$ and an upsampling operator to fuse them and match the spatial resolution of this level. 2) Addition of the skip image feature $\mathbf{f}_{S}$, followed by another channel attention block $\mathcal{N}_{CA}$ for joint refinement. 3) Addition of the metadata embedding $\mathbf{e}_{M}$, yielding the decoder output. This flow is expressed as:

$$\begin{cases} \mathbf{f}_{U}^{(k)} = Upsampling(\mathcal{N}_{CA}(Concate(\mathbf{f}_{DI}^{(k)}, \mathbf{e}_{U}^{(k)}))) \\ \mathbf{f}_{A}^{(k)} = \mathcal{N}_{CA}(\mathbf{f}_{U}^{(k)} + \mathbf{f}_{S}^{(k)}) \\ \mathbf{f}_{DO}^{(k)} = \mathbf{f}_{A}^{(k)} + \mathbf{e}_{M}^{(k)} \end{cases} . \tag{9}$$

By hierarchically integrating metadata awareness and undersampling prompts, our design empowers the decoder to progressively suppress artifacts and maintain high anatomical fidelity. The separation of encoder and decoder responsibilities promotes both generalizable representation learning and text-aware adaptive image reconstruction, thereby effectively modeling the underlying commonalities and heterogeneous characteristics of multimodal cardiovascular imaging.

## 2.4 Physics-informed data consistency module

In this module, each output is ensured to align with the acquired k-space data following the imaging physics (e.g., undersampling pattern and Fourier transform). Therefore, the physics-informed data consistency module is designed mostly same to the second sub-problem of (2) as follows:

$$\mathbf{x}^{(k)} = (\mathcal{F}^{*}\mathcal{U}^{*}\mathcal{U}\mathcal{F} + \lambda^{(k)})^{-1}(\mathcal{F}^{*}\mathcal{U}^{*}\mathbf{y} + \lambda^{(k)}\mathbf{m}^{(k)}), \tag{10}$$

and the only difference is that we set $\lambda$ as a trainable parameter initialized to 1. Specifically, (10) implies that, at the acquired positions, the data points should maintain a trade-off with $\mathbf{y}$, while the update of the non-acquired data points depends entirely on the network results.

In summary, in the proposed CardioMM, a text-aware image de-aliasing module followed by a physics-informed data consistency module constitutes a single network phase.

## 2.5 The tSNE visualization of text representations

Here, we performed t-SNE visualizations[22] on our CardioMM's text representations to investigate how the model organizes semantic priors derived from textual inputs. Specifically, we extracted representations from the metadata texts (focusing on imaging modality and field strength) and the undersampling texts (focusing on undersampling pattern and AF), after text encoder and projection heads. The goal of this analysis is to reveal whether CardioMM transforms explicit textual priors into a structured and continuous semantic manifold, such that it can retrieve semantically nearest information and generate meaningful conditioning for unseen combinations of data and text.

Supplementary Fig. 2a shows the t-SNE of metadata representations. Each point corresponds to a textual description of metadata; colors denote imaging modalities (e.g., cine, LGE, T1/T2 weighted, T1/T2 mapping, perfusion), and marker shapes represent field strengths (0.55T, 1.5T, 3.0T, 5.0T). Distinct clusters are formed for different modalities (e.g., cine, LGE, mapping, and weighted sequences occupy separable regions), demonstrating that the model captures modality-level semantic relationships rather than merely memorizing text patterns. Within each modality, points with different field strengths are mixed yet maintain a certain degree of independence, indicating that the learned representation is relatively robust to scanner-related parameters and primarily encodes semantic features relevant to modality type. Furthermore, smooth transitions between neighboring modalities (e.g., between cine and T1/T2 weighted clusters) suggest that the learned space preserves semantic continuity, allowing interpolation between related acquisition types. This continuous geometry allows the model to locate semantically meaningful neighbors when facing unseen metadata combinations, providing a basis for cross-modality generalization.

Supplementary Fig. 2b illustrates the t-SNE of undersampling representations, which reflects how the model organizes textual priors describing sampling geometry. Colors indicate undersampling patterns (uniform, random, radial), and marker shapes represent acceleration factor ranges (4×–8×, 8×–16×, 16×–24×). The three undersampling patterns form distinct, compact clusters, showing that the model effectively disentangles geometric semantics of different undersampling strategies. Within each cluster, AF levels are arranged in an orderly gradient, from lower (4×–8×) to higher (16×–24×), implying that the representation encodes continuous sensitivity to undersampling sparsity, rather than treating AF as a discrete categorical label. Notably, inter-pattern distances remain moderate rather than isolated, reflecting a semantically continuous manifold where different patterns maintain contextual proximity. This structure enables the text encoder to locate semantically closest regions and expand within their neighborhood when encountering unseen undersampling combinations, thereby exhibiting dynamic adaptability.

Together, these two visualizations demonstrate that CardioMM's text representation transforms explicit priors into a structured, hierarchical, and continuous semantic space. It disentangles major acquisition factors (modality and pattern) while maintaining smooth transitions across quantitative dimensions (field strength and AF). Consequently, when presented with unseen configurations, CardioMM can retrieve and extrapolate meaningful conditioning from neighboring regions in this semantic manifold, thereby enabling generalization and dynamic adaptation across diverse and unseen imaging scenarios.

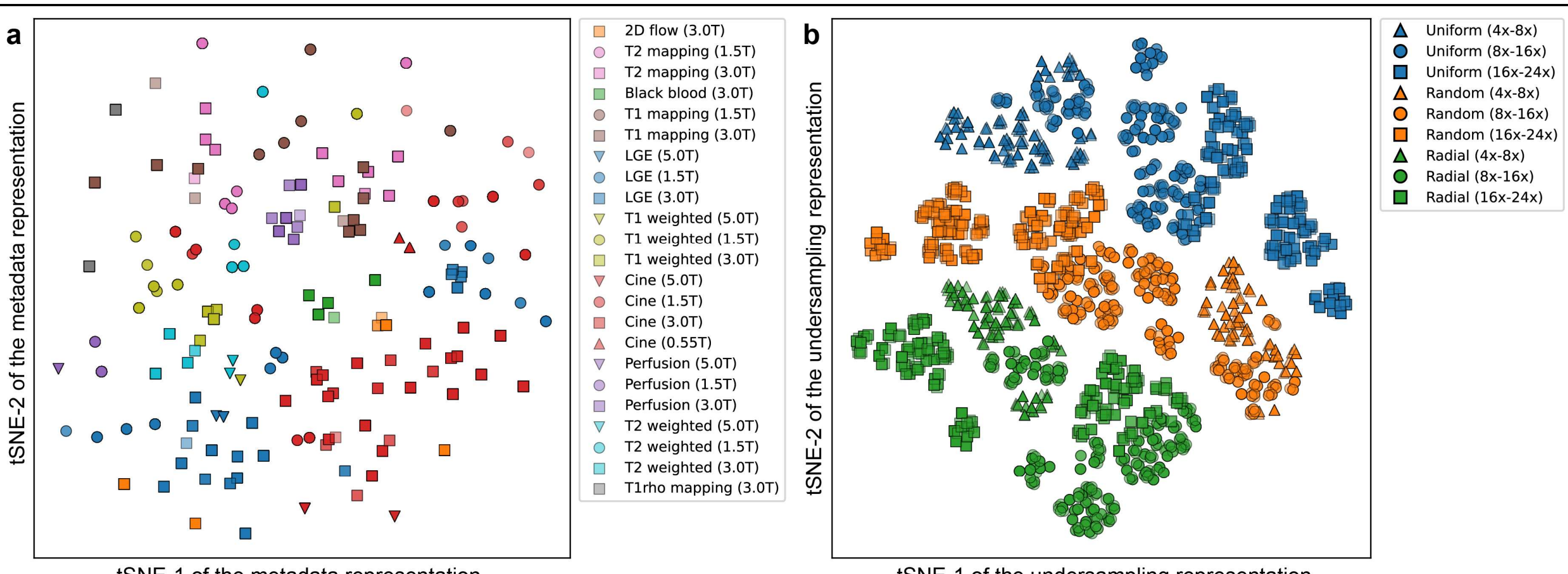

**Supplementary Fig. 2 | The tSNE visualization of CardioMM's text representations. a**, the latent feature distribution of the metadata representations. **b**, the latent feature distribution of the undersampling representations.

## Supplementary Note 3. More results of universal reconstruction across internal scenarios

**Supplementary Table 3 | Quantitative evaluation across eight internal centers, using three undersampling patterns (uniform, random, radial) with varying AFs (8×–24×) [Mean (95% CI)].**

| Method | PSNR (dB) | SSIM (%) |
|---|---|---|
| Conventional | 31.75 (31.68–31.82) * | 82.63 (82.48–82.78) * |
| DCUNet | 33.81 (33.73–33.88) * | 87.46 (87.27–87.65) * |
| PromptMR | 37.15 (37.06–37.24) * | 94.03 (93.94–94.12) * |
| CardioSM | 37.26 (37.17–37.34) * | 94.27 (94.19–94.35) * |
| CardioMM | **37.94 (37.86–38.03)** | **94.83 (94.76–94.90)** |

Note: This assessment involves 75,753 multi-coil k-space data from 1,495 scans of 320 participants, covering 12 CMR modalities acquired on routine high-field scanners (1.5T and 3.0T). The mean values and 95% CIs are computed over all tested data, respectively. The highest PSNR and SSIM values are bold faced. "*" means the compared method has statistically significant differences ($p<0.05$) compared to our CardioMM under two-sided t-test.

## Supplementary Note 4. More results of generalization capability across external centers

**Supplementary Table 4 | Quantitative evaluation across four external centers, using three undersampling patterns (uniform, random, radial) with varying AFs (4×–24×) [Mean (95% CI)].**

| Center | Method | PSNR (dB) | SSIM (%) |
|---|---|---|---|
| SHQT | Conventional | 32.58 (32.49–32.66) * | 84.01 (83.83–84.19) * |
| | DCUNet | 34.50 (34.41–34.59) * | 88.30 (88.07–88.53) * |
| | PromptMR | 38.01 (37.89–38.13) * | 94.47 (94.34–94.59) * |
| | CardioSM | 38.24 (38.13–38.35) * | 94.82 (94.71–94.92) * |
| | CardioMM | **38.87 (38.76–38.98)** | **95.32 (95.23–95.41)** |
| SHSX | Conventional | 31.47 (31.25–31.68) * | 82.04 (81.50–82.57) * |
| | DCUNet | 31.33 (31.11–31.55) * | 84.34 (83.88–84.81) * |
| | PromptMR | 35.44 (35.20–35.68) * | 91.45 (91.13–91.76) * |
| | CardioSM | 35.53 (35.28–35.79) * | 91.36 (91.01–91.70) * |
| | CardioMM | **36.09 (35.83–36.36)** | **91.86 (91.53–92.19)** |
| EJHS | Conventional | 30.39 (29.69–31.09) * | 76.90 (74.23–79.56) * |
| | DCUNet | 30.67 (29.17–32.14) * | 75.33 (71.87–78.79) * |
| | PromptMR | 33.26 (31.47–35.03) * | 85.64 (81.94–89.34) * |
| | CardioSM | 33.99 (32.48–35.54) | 89.90 (87.70–92.12) |
| | CardioMM | **34.55 (33.16–35.91)** | **90.90 (89.06–92.73)** |
| UKSK | Conventional | 28.80 (28.69–28.91) * | 81.31 (81.02–81.60) * |
| | DCUNet | 26.42 (26.33–26.52) * | 75.38 (75.17–75.59) * |
| | PromptMR | 31.71 (31.58–31.84) * | 87.83 (87.65–88.02) * |
| | CardioSM | 32.00 (31.87–32.13) * | 88.36 (88.17–88.55) * |
| | CardioMM | **32.28 (32.15–32.42)** | **88.78 (88.58–88.97)** |

Note: This assessment involves 101,069 multi-coil k-space datasets from 1,115 scans of 321 participants, covering seven major CMR modalities acquired on routine high-field scanners (1.5T and 3.0T). The mean values and 95% CIs are computed over all tested data, respectively. The highest PSNR and SSIM values are bold faced. “*” means the compared method has statistically significant differences ($p<0.05$) compared to our CardioMM under Wilcoxon signed-rank test.

## Supplementary Note 5. More results of generalization capability across external field strengths

**Supplementary Table 5 | Quantitative evaluation across two external field strengths from three centers, using three undersampling patterns (uniform, random, radial) with varying AFs (8×–24×) [Mean (95% CI)].**

| System | Method | PSNR (dB) | SSIM (%) |
|---|---|---|---|
| 0.55T | Conventional | 30.04 (29.65–30.43) * | 73.05 (71.56–74.55) * |
| | DCUNet | 33.58 (33.16–34.01) * | 87.72 (86.39–89.06) * |
| | PromptMR | 35.80 (35.34–36.26) * | 90.18 (88.47–91.89) |
| | CardioSM | 36.00 (35.56–36.46) * | 90.42 (88.76–92.08) |
| | CardioMM | **36.40 (35.94–36.86)** | **90.70 (88.99–92.41)** |
| 5.0T | Conventional | 33.36 (33.13–33.60) * | 84.32 (83.80–84.84) * |
| | DCUNet | 35.22 (34.96–35.48) * | 89.35 (88.87–89.82) * |
| | PromptMR | 38.34 (38.04–38.64) * | 94.63 (94.06–95.21) * |
| | CardioSM | 38.40 (38.10–38.70) * | 94.81 (94.25–95.38) |
| | CardioMM | **38.91 (38.60–39.22)** | **95.12 (94.56–95.68)** |

Note: This assessment involves 9,117 multi-coil k-space datasets from 110 scans of 74 participants, covering five major CMR modalities acquired on low-field and ultra-high-filed scanners (0.55T and 5.0T). The mean values and 95% CIs are computed over all tested data, respectively. The highest PSNR and SSIM values are bold faced. “*” means the compared method has statistically significant differences ($p<0.05$) compared to our CardioMM under Wilcoxon signed-rank test.

# Supplementary Note 6. More results of clinical applicability of automated imaging phenotyping for diagnostic support

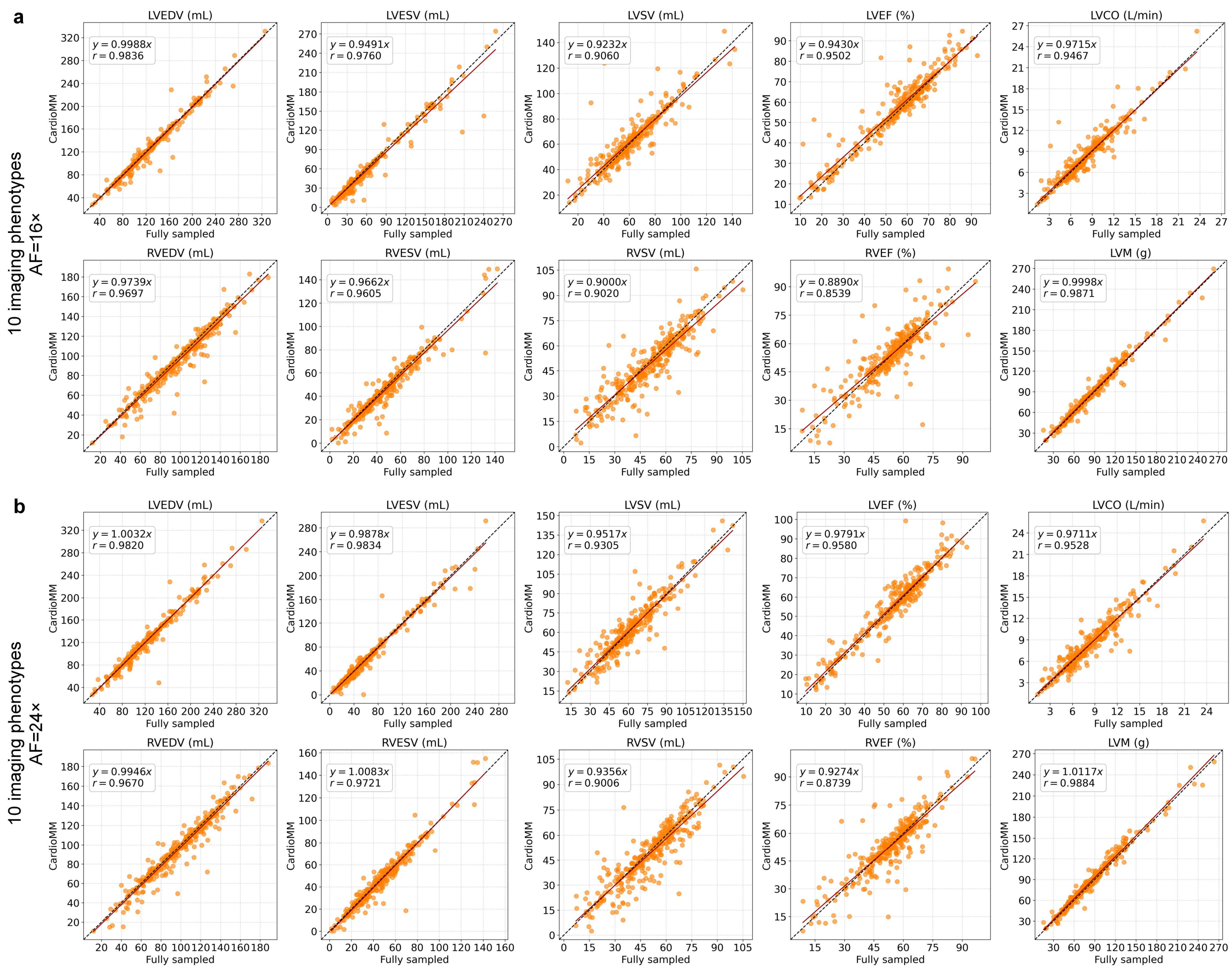

**Supplementary Fig. 3 | Correlation analysis of 10 representative cardiac imaging phenotypes derived from fully sampled and CardioMM-reconstructed images. a,** linear regression and PCC analysis at AF=16× for each phenotype. **b,** linear regression and PCC analysis at AF=24× for each phenotype. Note: *r* corresponds to the PCC. This assessment involves 355 participants with multi-slice short-axis cine modality. Based on previous study about the suitability of different undersampling patterns at varying AFs[23], these undersampling settings (random AF=16×, radial AF=24×) are adopted here to enable higher accelerations.

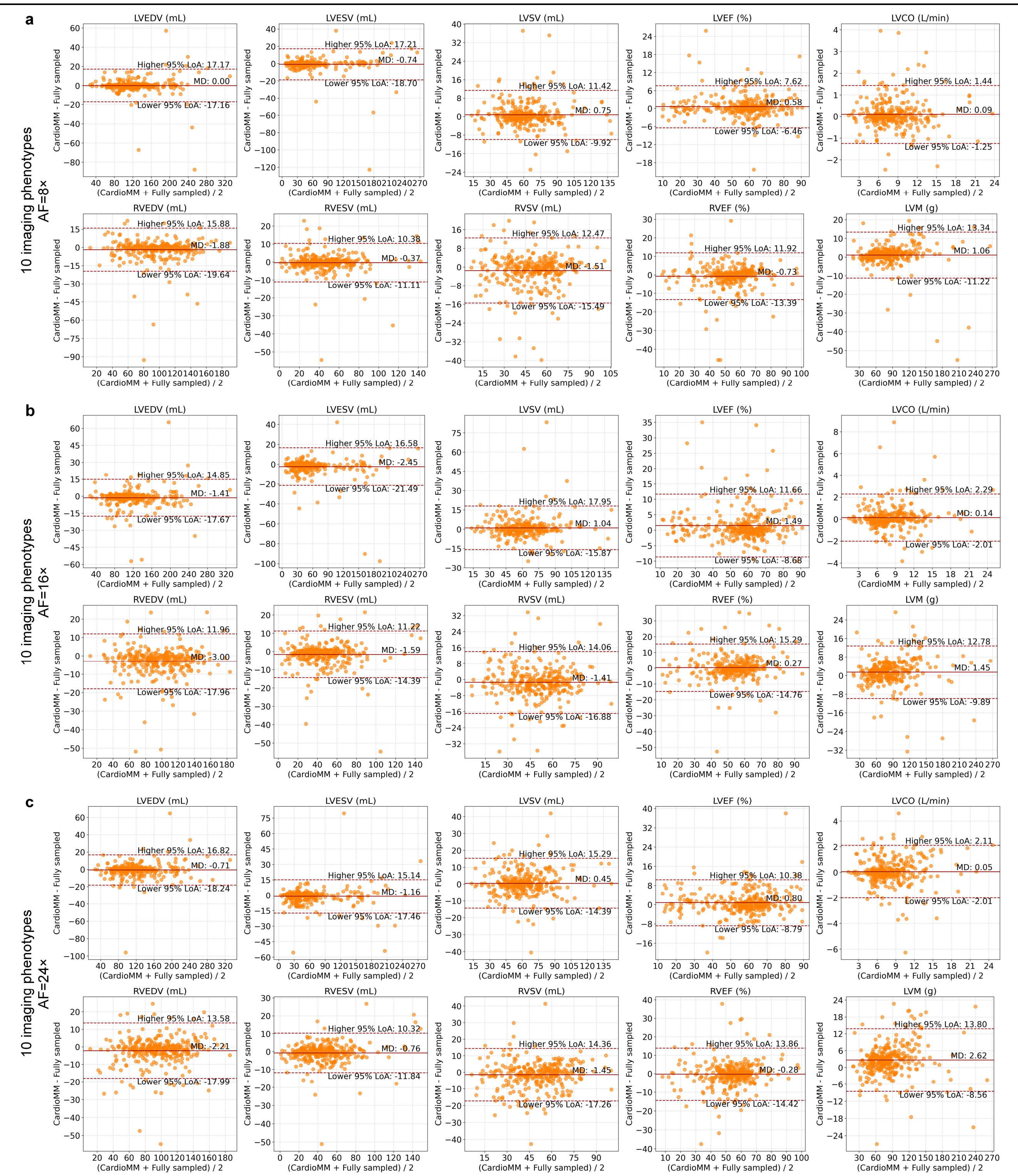

**Supplementary Fig. 4 | Bland-Altman analysis of 10 representative cardiac imaging phenotypes derived from fully sampled and CardioMM-reconstructed images. a,** Bland-Altman analysis at AF=8× for each phenotype. **b,** Bland-Altman analysis at AF=16× for each phenotype. **c,** Bland-Altman analysis at AF=24× for each phenotype. Note: "MD" is the mean difference, and "LoA" is the limits of agreement. This assessment involves 355 participants with multi-slice short-axis cine modality. Based on previous study about the suitability of different undersampling patterns at varying AFs[23], these undersampling settings (uniform AF=8×, random AF=16×, radial AF=24×) are adopted here to enable higher accelerations.

**Supplementary Table 6 | PCCs (*r*) of 10 representative cardiac imaging phenotypes derived from fully sampled and reconstructed images obtained by different methods.**

| AF | Method | LVEDV | LVESV | LVSV | LVEF | LVCO | LVM | RVEDV | RVESV | RVSV | RVEF |
|---|---|---|---|---|---|---|---|---|---|---|---|
| 8× | Conventional | 0.7565 | 0.7237 | 0.6542 | 0.6018 | 0.7356 | 0.7208 | 0.4727 | 0.4340 | 0.5580 | 0.4784 |
| | DCUNet | 0.9287 | 0.9358 | 0.8169 | 0.8758 | 0.8812 | 0.9403 | 0.8907 | 0.9170 | 0.8011 | 0.7803 |
| | PromptMR | 0.9532 | 0.8915 | 0.9477 | 0.9668 | 0.9701 | 0.9685 | 0.9440 | 0.9607 | 0.8956 | 0.8729 |
| | CardioSM | 0.9682 | 0.9675 | 0.9568 | 0.9646 | 0.9712 | 0.9748 | 0.9504 | **0.9742** | 0.9094 | **0.9058** |
| | CardioMM | **0.9820** | **0.9795** | **0.9629** | **0.9767** | **0.9794** | **0.9850** | **0.9569** | 0.9722 | **0.9205** | 0.8981 |
| 16× | Conventional | 0.7725 | 0.5329 | 0.3629 | 0.4016 | 0.6044 | 0.8142 | 0.5744 | 0.2260 | 0.4762 | 0.0879 |
| | DCUNet | 0.9328 | 0.8996 | 0.8036 | 0.8359 | 0.8720 | 0.9311 | 0.8378 | 0.7973 | 0.7328 | 0.6102 |
| | PromptMR | **0.9864** | 0.9697 | 0.8986 | 0.9401 | 0.9461 | 0.9855 | 0.9596 | 0.9424 | **0.9088** | 0.8537 |
| | CardioSM | 0.9810 | **0.9773** | **0.9081** | 0.9347 | 0.9461 | 0.9862 | 0.9627 | 0.9548 | 0.9079 | 0.8506 |
| | CardioMM | 0.9836 | 0.9760 | 0.9060 | **0.9502** | **0.9467** | **0.9871** | **0.9697** | **0.9605** | 0.9020 | **0.8539** |
| 24× | Conventional | 0.7678 | 0.7135 | 0.5876 | 0.5717 | 0.6809 | 0.7447 | 0.6399 | 0.5976 | 0.6302 | 0.5132 |
| | DCUNet | 0.9376 | 0.8915 | 0.7341 | 0.7920 | 0.8277 | 0.9154 | 0.8601 | 0.8168 | 0.7917 | 0.6963 |
| | PromptMR | 0.9753 | 0.9739 | 0.9212 | 0.9482 | 0.9469 | 0.9785 | 0.9417 | 0.9545 | 0.8724 | 0.8590 |
| | CardioSM | 0.9750 | 0.9679 | 0.8965 | 0.9408 | 0.9341 | 0.9800 | 0.9529 | 0.9574 | 0.8929 | 0.8426 |
| | CardioMM | **0.9820** | **0.9834** | **0.9305** | **0.9580** | **0.9528** | **0.9884** | **0.9670** | **0.9721** | **0.9006** | **0.8739** |

Note: This assessment involves 355 participants with multi-slice short-axis cine modality. Based on previous study about the suitability of different undersampling patterns at varying AFs[23], these undersampling settings (uniform AF=8×, random AF=16×, radial AF=24×) are adopted here to enable higher accelerations. The highest PCCs (*r*) are bold faced.

**Supplementary Table 7 | Bland-Altman analysis results of 10 representative cardiac imaging phenotypes derived from fully sampled and reconstructed images obtained by different methods [Mean difference (95% LoA)].**

| AF | Method | LVEDV (mL) | LVESV (mL) | LVSV (mL) | LVEF (%) | LVCO (L/min) | LVM (g) | RVEDV (mL) | RVESV (mL) | RVSV (mL) | RVEF (%) |
|---|---|---|---|---|---|---|---|---|---|---|---|
| 8× | Conventional | -23.66<br>(-74.97, 27.64) | -26.89<br>(-80.15, 26.37) | 3.22<br>(-28.20, 34.65) | 16.15<br>(-9.93, 42.24) | 0.35<br>(-3.86, 4.56) | -21.37<br>(-64.09, 21.35) | -35.96<br>(-93.60, 21.68) | -25.33<br>(-63.38, 12.71) | 11.56<br>(-22.68, 45.80) | -10.62<br>(-46.31, 25.06) |
| | DCUNet | -8.02<br>(-40.32, 24.29) | -7.36<br>(-37.27, 22.54) | -0.65<br>(-24.13, 22.82) | 3.20<br>(-13.13, 19.53) | -0.08<br>(-3.25, 3.10) | -4.20<br>(-27.14, 18.74) | -11.04<br>(-39.04, -11.04) | -5.50<br>(-23.39, 12.40) | -5.54<br>(-27.69, 16,61) | 0.47<br>(-19.23, 20.17) |
| | PromptMR | -1.74<br>(-29.12, 25.64) | -1.94<br>(-28.22, 24.34) | **0.19**<br>**(-12.37, 12.76)** | 0.81<br>(-7.65, 9.26) | **0.02**<br>**(-1.58, 1.63)** | **0.81**<br>**(-16.93, 18.56)** | -3.38<br>(-23.75, 16.99) | -1.51<br>(-14.24, 11.21) | -1.87<br>(-17.92, 14.18) | -0.30<br>(-14.75, 14.16) |
| | CardioSM | -1.70<br>(-24.38, 20.98) | -2.13<br>(-24.70, 20.45) | 0.43<br>(-10.82, 11.68) | 1.00<br>(-7.65, 9.65) | 0.06<br>(-1.49, 1.61) | 1.07<br>(-14.86, 17.00) | -3.32<br>(-22.33, 15.68) | -1.54<br>(-11.80, 8.73) | -1.79<br>(-16.64, 13.07) | **-0.06**<br>**(-12.11, 11.99)** |
| | CardioMM | **0.00**<br>**(-17.16, 17.17)** | **-0.74**<br>**(-18.70, 17.21)** | 0.75<br>(-9.92, 11.42) | **0.58**<br>**(-6.46, 7.62)** | 0.09<br>(-1.25, 1.44) | 1.06<br>(-11.22, 13.34) | **-1.88**<br>**(-19.64, 15.88)** | **-0.37**<br>**(-11.11, 10.38)** | **-1.51**<br>**(-15.49, 12.47)** | -0.73<br>(-13.39, 11.92) |
| 16× | Conventional | -24.18<br>(-75.30, 26.93) | -33.08<br>(-103.02, 36.86) | 8.90<br>(-36.93, 54.72) | 20.56<br>(-10.61, 51.74) | 1.12<br>(-4.70, 6.94) | -22.12<br>(-61.79, 17.56) | -40.16<br>(-89.06, 8.75) | -32.01<br>(-77.88, 13.86) | -8.15<br>(-47.15, 30.86) | 17.90<br>(-24.30, 60.11) |
| | DCUNet | -8.84<br>(-40.41, 22.74) | -10.14<br>(-47.43, 27.16) | 1.30<br>(-23.99, 26.59) | 5.15<br>(-13.07, 23.37) | 0.19<br>(-3.27, 3.65) | -4.88<br>(-29.80, 20.04) | 14.09<br>(-47.72, 19.54) | -7.69<br>(-34.88, 19.49) | -6.40<br>(-31.87, 19.08) | 1.29<br>(-24.08, 26.66) |
| | PromptMR | -2.54<br>(-17.07, 12.00) | -3.80<br>(-25.21, 17.62) | 1.26<br>(-16.26, 18.78) | 2.13<br>(9.07, 13.32) | 0.16<br>(-2.00, 2.31) | **1.25**<br>**(-10.51, 13.02)** | -4.69<br>(-21.71, 12.33) | -2.92<br>(-18.23, 12.38) | -1.77<br>(-16.51, 12.97) | 1.00<br>(-14.24, 16.23) |
| | CardioSM | -2.75<br>(-19.96, 14.47) | -4.21<br>(-22.90, 14.47) | 1.47<br>(-15.16, 18.10) | 2.54<br>(-9.12, 14.20) | 0.22<br>(-1.99, 2.42) | 2.29<br>(-9.29, 13.87) | -5.08<br>(-21.67, 11.50) | -3.05<br>(-16.56, 10.46) | -2.04<br>(-17.17, 13.10) | 0.76<br>(-14.42, 15.94) |
| | CardioMM | **-1.41**<br>**(-17.67, 14.85)** | **-2.45**<br>**(-21.49, 16.58)** | **1.04**<br>**(-15.87, 17.95)** | **1.49**<br>**(-8.68, 11.66)** | **0.14**<br>**(-2.01, 2.29)** | 1.45<br>(-9.89, 12.78) | **-3.00**<br>**(-17.96, 11.96)** | **-1.59**<br>**(-14.39, 11.22)** | **-1.41**<br>**(-16.88, 14.06)** | **0.27**<br>**(-14.76, 15.29)** |
| 24× | Conventional | -25.25<br>(-83.90, 33.39) | -31.29<br>(-96.25, 33.66) | 6.04<br>(-28.80, 40.87) | 17.89<br>(-10.77, 46.72) | 0.74<br>(-4.12, 5.59) | -20.34<br>(-65.49, 24.81) | -34.76<br>(-82.37, 12.84) | -24.39<br>(-58.84, 10.07) | -10.38<br>(-42.97, 22,21) | 10.21<br>(-22.98, 43.41) |
| | DCUNet | -7.18<br>(-38.11, 23.76) | -11.37<br>(-51.52, 28.77) | 4.20<br>(-22.83, 31.22) | 6.60<br>(-13.97, 27.18) | 0.61<br>(-3.21, 4.42) | -3.82<br>(-29.64, 22.00) | 12.22<br>(-43.39, 18.96) | -8.39<br>(-35.27, 18.49) | -3.83<br>(-27.51, 19.85) | 2.88<br>(-21.78, 27.54) |
| | PromptMR | -3.29<br>(-22.70, 16.12) | -4.58<br>(-24.19, 15.02) | 1.29<br>(-14.38, 16.97) | 2.59<br>(-7.88, 13.06) | 0.18<br>(-2.03, 2.38) | **0.77**<br>**(-13.63, 15.17)** | -5.17<br>(-25.60, 15.26) | -3.59<br>(-16.91, 9.74) | -1.59<br>(-19.72, 16.55) | 1.36<br>(-13.94, 16.66) |
| | CardioSM | -2.53<br>(-22.65, 17.59) | -3.41<br>(-25.62, 18.80) | 0.88<br>(-16.93, 18.68) | 1.87<br>(-9.42, 13.16) | 0.13<br>(-2.30, 2.57) | 2.23<br>(-12.11, 16.58) | -4.04<br>(-22.84, 14.75) | -2.19<br>(-15.25, 10.88) | -1.85<br>(-18.25, 14.54) | 0.17<br>(-16.09, 16.44) |
| | CardioMM | **-0.71**<br>**(-18.24, 16.82)** | **-1.16**<br>**(-17.46, 15.14)** | **0.45**<br>**(-14.39, 15.29)** | **0.80**<br>**(-8.79, 10.38)** | **0.05**<br>**(-2.01, 2.11)** | 2.62<br>(-8.56, 13.80) | **-2.21**<br>**(-17.99, 13.58)** | **-0.76**<br>**(-11.84, 10.32)** | **-1.45**<br>**(-17.26, 14.36)** | **-0.28**<br>**(-14.42, 13.86)** |

Note: This assessment involves 355 participants with multi-slice short-axis cine modality. Based on previous study about the suitability of different undersampling patterns at varying AFs[23], these undersampling settings (uniform AF=8×, random AF=16×, radial AF=24×) are adopted here to enable higher accelerations. The lowest absolute mean differences are bold faced.

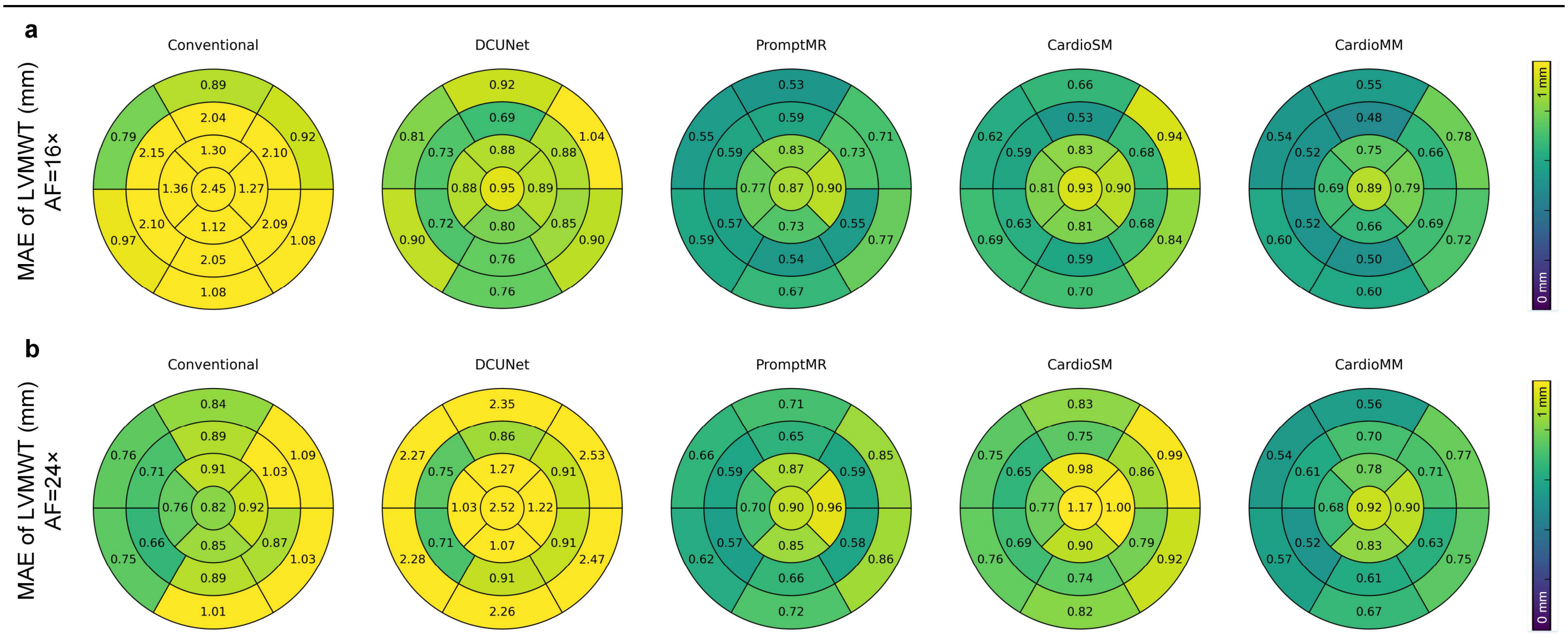

**Supplementary Fig. 5 | Average MAE of LVMWT between fully sampled reference and different methods based on the AHA 16-segment model with a global segment. a,** Bullseye charts at AF=16×. **b,** Bullseye charts at AF=24×. Note: This assessment involves 355 participants with multi-slice short-axis cine modality. Based on previous study about the suitability of different undersampling patterns at varying AFs[23], these undersampling settings (random AF=16×, radial AF=24×) are adopted here to enable higher accelerations.

**Supplementary Table 8 | Diagnostic performance (AUCs) of three cardiac phenotypes derived from fully sampled and reconstructed images obtained by different methods.**

| AF | Method | LVEDV-based DCM diagnosis | LVEF-based HF diagnosis | LVMWT-based HCM diagnosis |
|---|---|---|---|---|
| 1× | Fully sampled | 0.9633 | 0.9771 | 0.9806 |
| 8× | Conventional | 0.4664 * | 0.4342 * | 0.5298 * |
| | DCUNet | 0.9000 * | 0.8533 * | 0.8499 * |
| | PromptMR | 0.9518 | 0.9413 | 0.9265 * |
| | CardioSM | **0.9640** | 0.9344 * | 0.9073 * |
| | CardioMM | **0.9640** | **0.9569** | **0.9811** |
| 16× | Conventional | 0.4834 * | 0.3327 * | 0.4933 * |
| | DCUNet | 0.8876 * | 0.7728 * | 0.8406 * |
| | PromptMR | 0.9419 | 0.8884 * | 0.9300 * |
| | CardioSM | 0.9453 | 0.8739 * | 0.9243 * |
| | CardioMM | **0.9490** | **0.9360** | **0.9845** |
| 24× | Conventional | 0.5823 * | 0.5044 * | 0.6112 * |
| | DCUNet | 0.8863 * | 0.7645 * | 0.7620 * |
| | PromptMR | 0.9203 * | 0.9124 * | 0.9003 * |
| | CardioSM | **0.9447** | 0.8984 * | 0.9120 * |
| | CardioMM | 0.9380 | **0.9531** | **0.9768** |

Note: This assessment involves 122 participants (52 DCM patients and 70 healthy controls) for DCM diagnosis; 149 participants (79 HF patients and 70 healthy controls) for HF diagnosis; 150 participants (80 HCM patients and 70 healthy controls) for HCM diagnosis. Based on previous study about the suitability of different undersampling patterns at varying AFs[23], these undersampling settings (uniform AF=8×, random AF=16×, radial AF=24×) are adopted here to enable higher accelerations. The highest AUC values of reconstruction methods are bold faced. "*" means the reconstruction method has statistically significant differences ($p<0.05$) compared to fully sampled reference under Bootstrap resampling test.

# Supplementary Note 7. More results of clinical applicability of quantitative myocardial biomarkers for diagnostic support

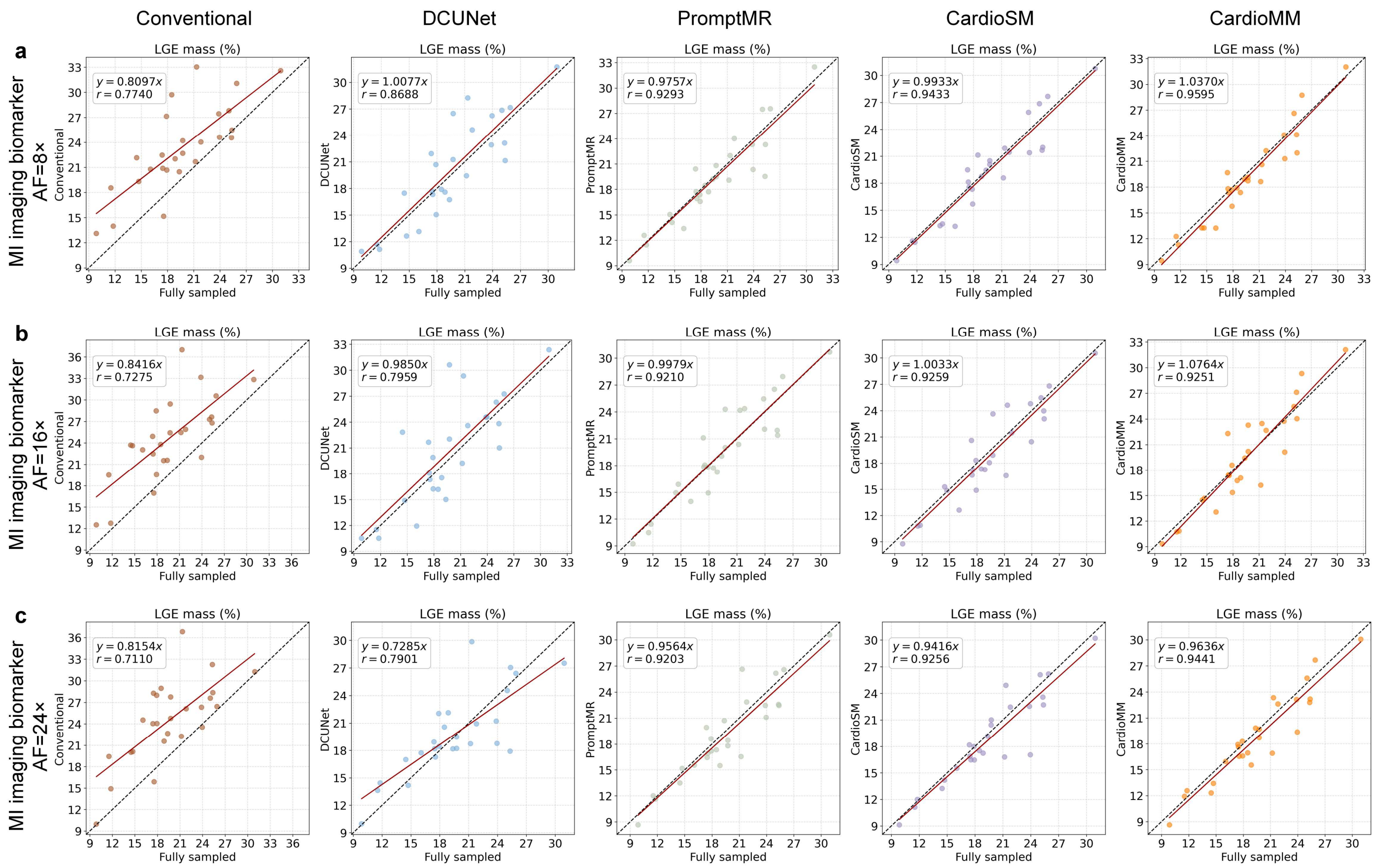

**Supplementary Fig. 6 | Correlation analysis of the myocardial infarction (MI) imaging biomarker (LGE mass) derived from fully sampled and reconstructed images obtained by different methods. a,** linear regression and PCC analysis at AF=8×. **b,** linear regression and PCC analysis at AF=16×. **c,** linear regression and PCC analysis at AF=24×. Note: *r* corresponds to the PCC. This assessment involves 26 MI patients with multi-slice short-axis LGE modality. Based on previous study about the suitability of different undersampling patterns at varying AFs[23], these undersampling settings (random AF=16×, radial AF=24×) are adopted here to enable higher accelerations.

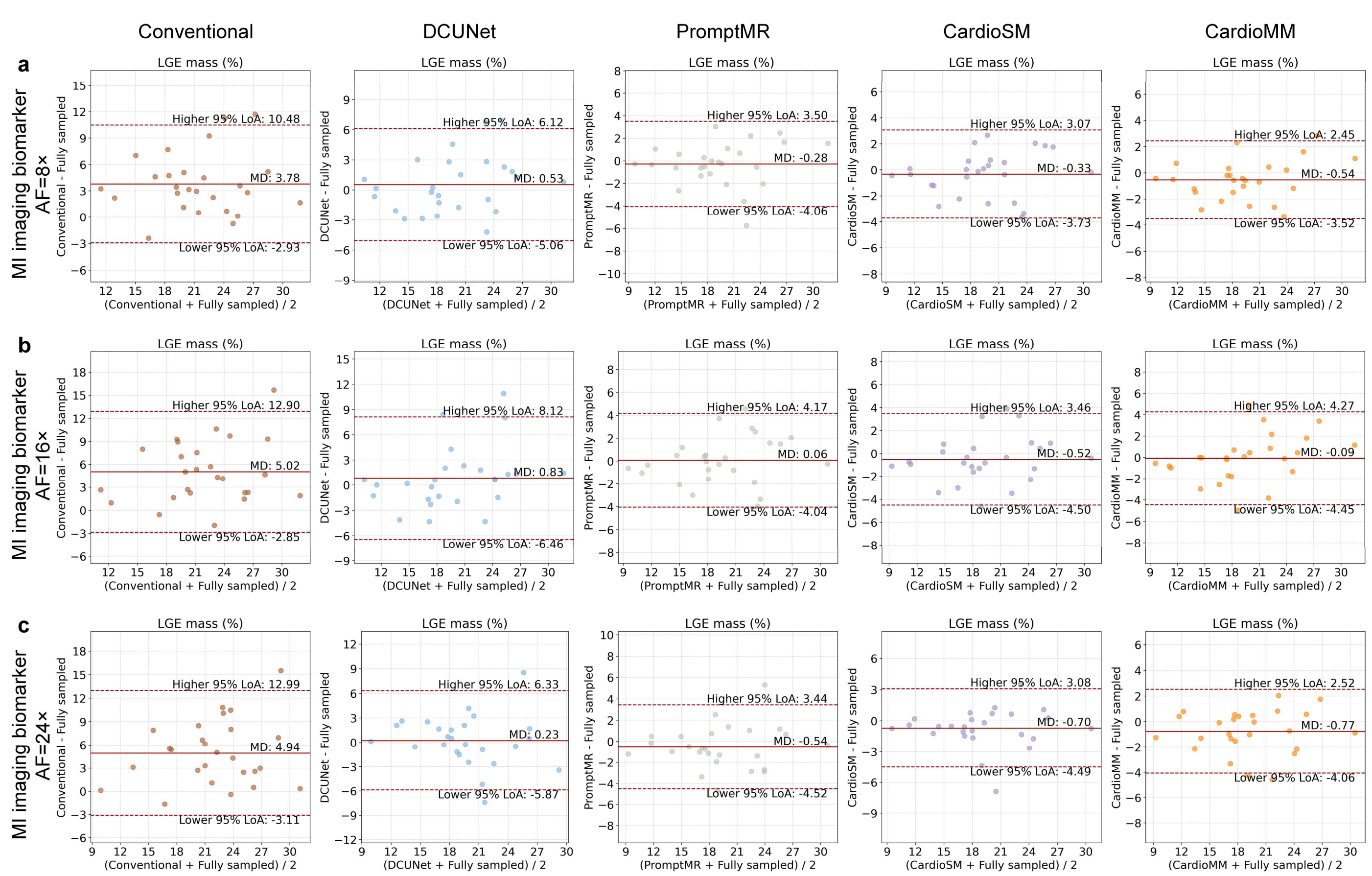

**Supplementary Fig. 7 | Bland-Altman analysis of the myocardial infarction (MI) imaging biomarker (LGE mass) derived from fully sampled and reconstructed images obtained by different methods. a,** Bland-Altman analysis at AF=8×. **b,** Bland-Altman analysis at AF=16×. **c,** Bland-Altman analysis at AF=24×. Note: "MD" is the mean difference, and "LoA" is the limits of agreement. This assessment involves 26 MI patients with multi-slice short-axis LGE modality. Based on previous study about the suitability of different undersampling patterns at varying AFs[23], these undersampling settings (uniform AF=8×, random AF=16×, radial AF=24×) are adopted here to enable higher accelerations.

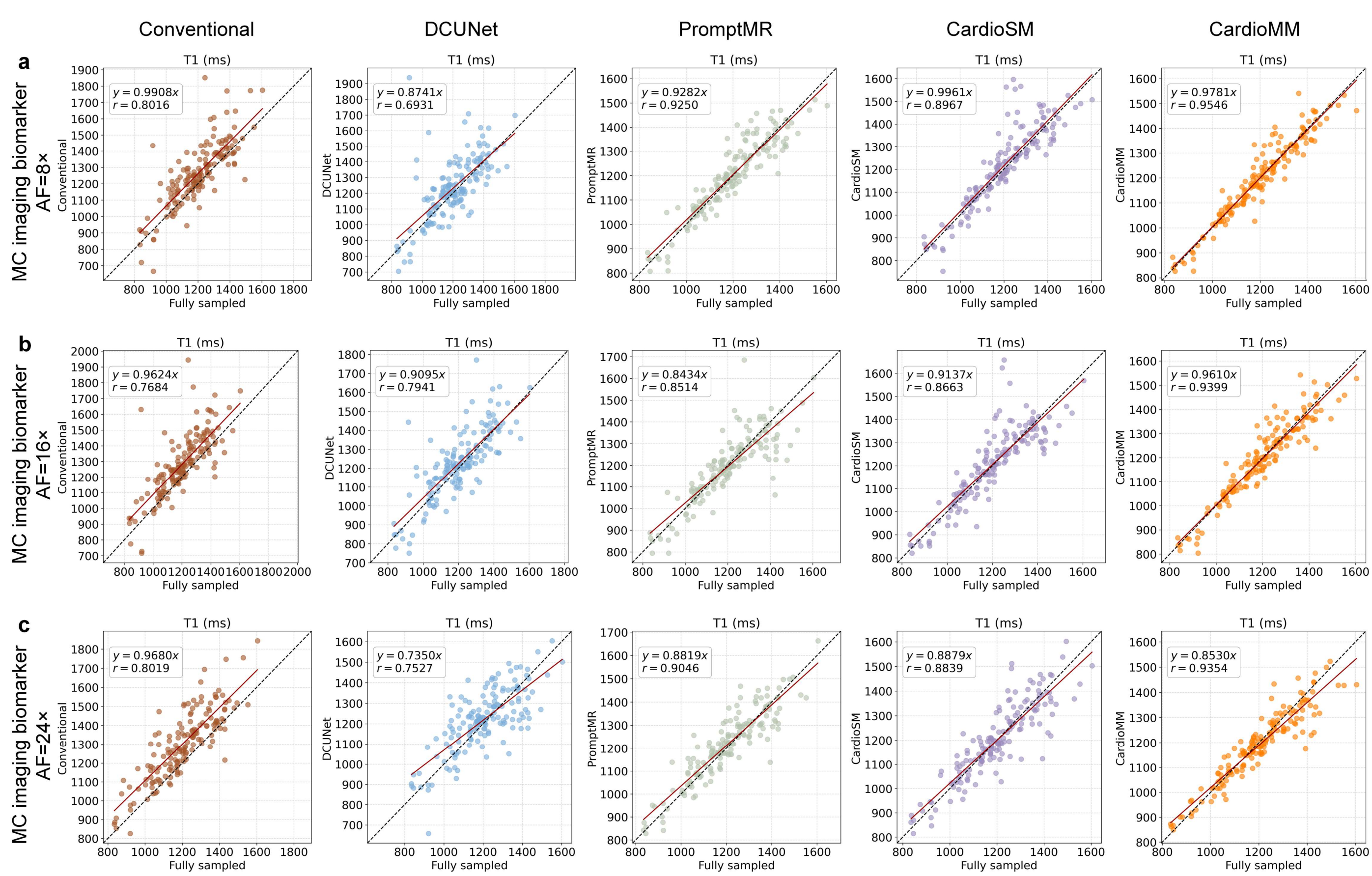

**Supplementary Fig. 8 | Correlation analysis of the myocarditis (MC) imaging biomarker (T1) derived from fully sampled and reconstructed images obtained by different methods. a,** linear regression and PCC analysis at AF=8×. **b,** linear regression and PCC analysis at AF=16×. **c,** linear regression and PCC analysis at AF=24×. Note: *r* corresponds to the PCC. This assessment involves 10 MC patients with multi-slice short-axis T1 mapping modality, and each dot represents a segment-wise T1 value from the AHA 16-segment model. Based on previous study about the suitability of different undersampling patterns at varying AFs[23], these undersampling settings (random AF=16×, radial AF=24×) are adopted here to enable higher accelerations.

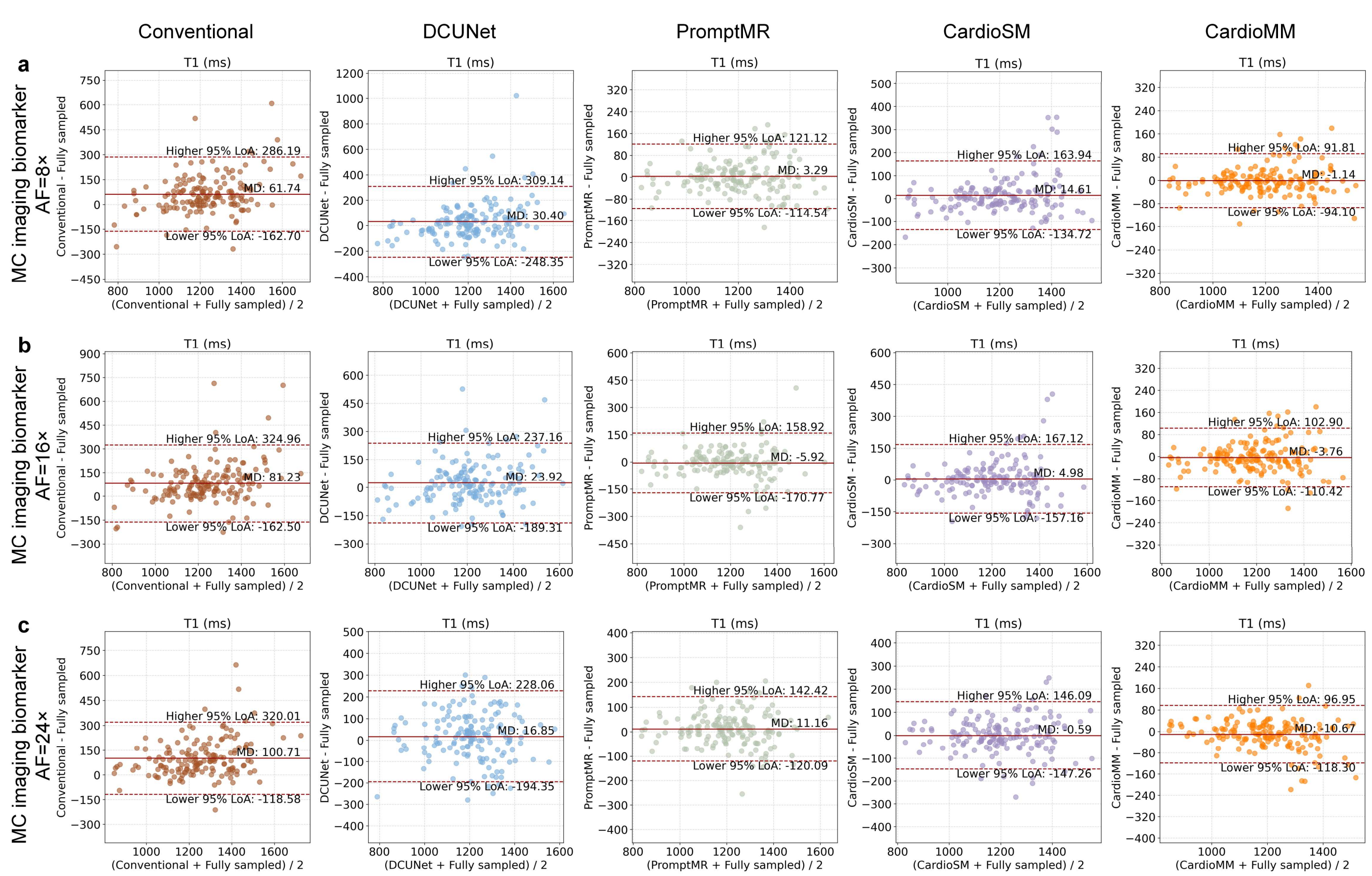

**Supplementary Fig. 9 | Bland-Altman analysis of the myocarditis (MC) imaging biomarker (T1) derived from fully sampled and reconstructed images obtained by different methods. a,** Bland-Altman analysis at AF=8×. **b,** Bland-Altman analysis at AF=16×. **c,** Bland-Altman analysis at AF=24×. Note: "MD" is the mean difference, and "LoA" is the limits of agreement. This assessment involves 10 MC patients with multi-slice short-axis T1 mapping modality, and each dot represents a segment-wise T1 value from the AHA 16-segment model. Based on previous study about the suitability of different undersampling patterns at varying AFs[23], these undersampling settings (uniform AF=8×, random AF=16×, radial AF=24×) are adopted here to enable higher accelerations.

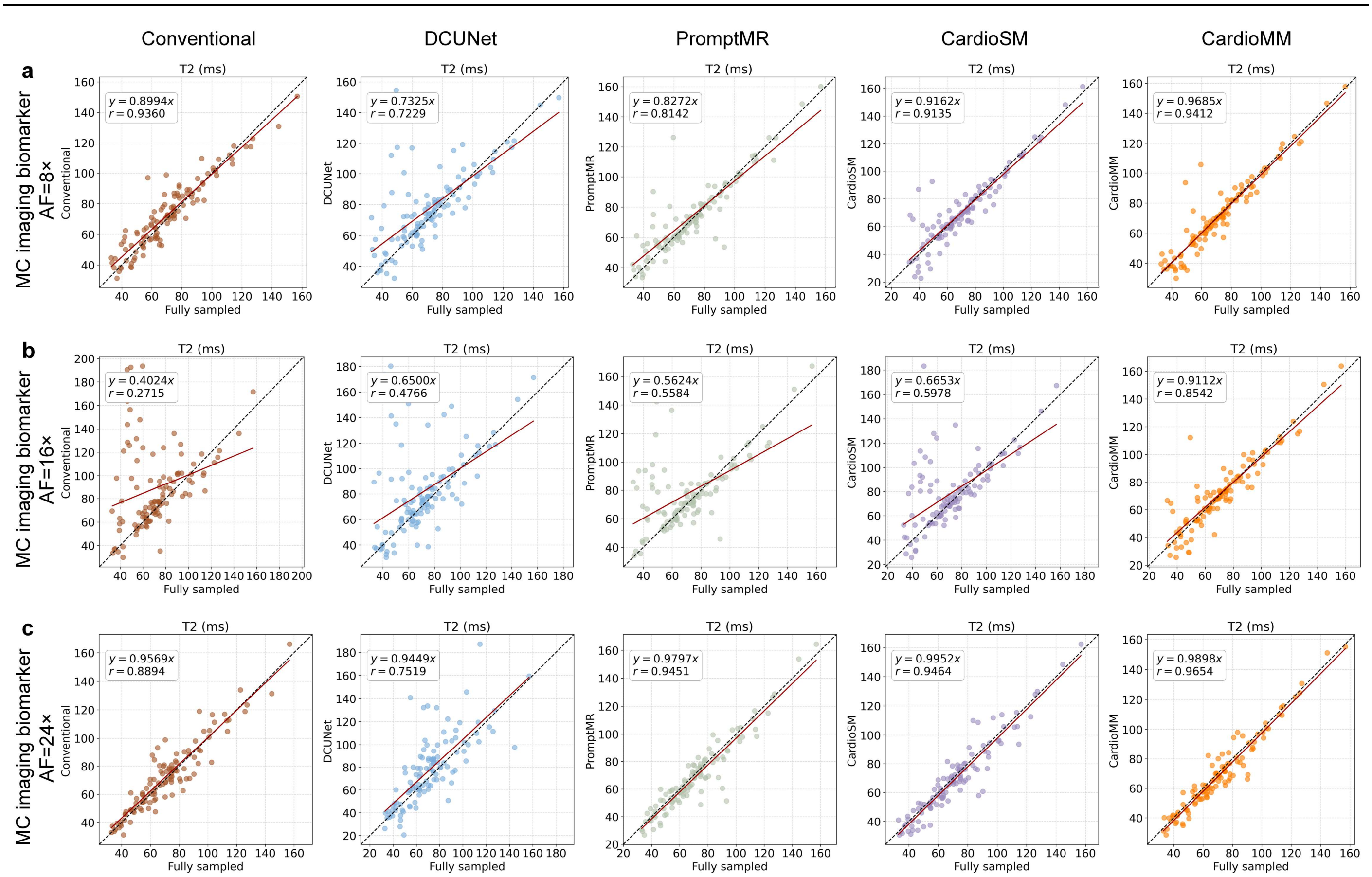

**Supplementary Fig. 10 | Correlation analysis of the myocarditis (MC) imaging biomarker (T2) derived from fully sampled and reconstructed images obtained by different methods. a,** linear regression and PCC analysis at AF=8×. **b,** linear regression and PCC analysis at AF=16×. **c,** linear regression and PCC analysis at AF=24×. Note: $r$ corresponds to the PCC. This assessment involves 10 MC patients with multi-slice short-axis T2 mapping modality, and each dot represents a segment-wise T2 value from the AHA 16-segment model. Based on previous study about the suitability of different undersampling patterns at varying AFs[23], these undersampling settings (random AF=16×, radial AF=24×) are adopted here to enable higher accelerations.

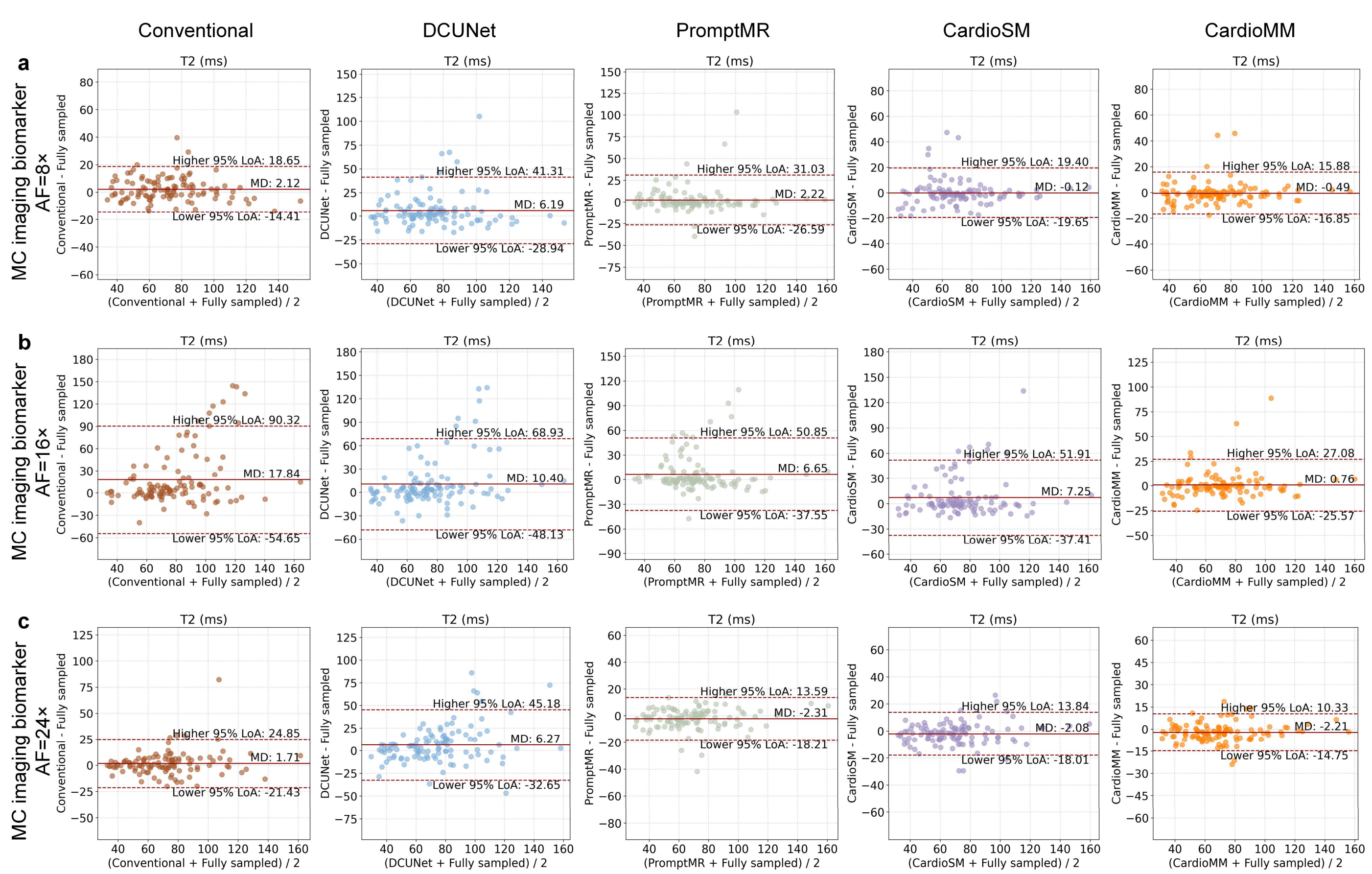

**Supplementary Fig. 11 | Bland-Altman analysis of the myocarditis (MC) imaging biomarker (T2) derived from fully sampled and reconstructed images obtained by different methods. a,** Bland-Altman analysis at AF=8×. **b,** Bland-Altman analysis at AF=16×. **c,** Bland-Altman analysis at AF=24×. Note: "MD" is the mean difference, and "LoA" is the limits of agreement. This assessment involves 10 MC patients with multi-slice short-axis T2 mapping modality, and each dot represents a segment-wise T2 value from the AHA 16-segment model. Based on previous study about the suitability of different undersampling patterns at varying AFs[23], these undersampling settings (uniform AF=8×, random AF=16×, radial AF=24×) are adopted here to enable higher accelerations.

## Supplementary Note 8. Results of reader study for qualitative assessment

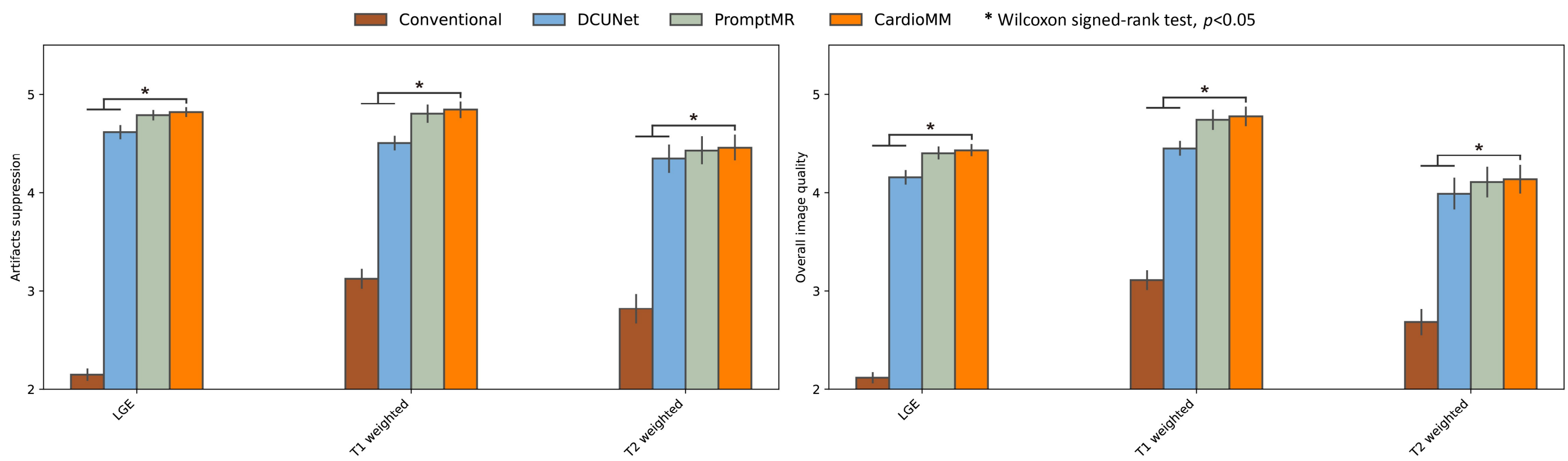

**Supplementary Fig. 12 | Reader study.** Scores of fully sampled references and different reconstructed images are shown for each scored modality. Note: This evaluation is conducted across seven centers, using three undersampling patterns (uniform, random, radial) with varying AFs (8×–24×). The reported mean values and 95% CIs in the bar charts are computed over all tested data for each modality, respectively.

**Supplementary Table 9 | Reader study across seven centers, using three undersampling patterns (uniform, random, radial) with varying AFs (8×–24×) [Mean (95% CI)].**

| Method | Artifacts suppression | Overall image quality |
| --- | --- | --- |
| Conventional | 2.47 (2.41–2.54) * | 2.42 (2.36–2.48) * |
| DCUNet | 4.54 (4.48–4.59) * | 4.17 (4.11–4.23) * |
| PromptMR | 4.71 (4.66–4.76) | 4.40 (4.34–4.46) |
| CardioMM | **4.74 (4.70–4.79)** | **4.43 (4.37–4.49)** |

Note: This assessment involves 168 participants with 103 LGE scans, 73 T1 weighted scans, and 88 T2 weighted scans, acquired on routine high-field scanners (1.5T and 3.0T). The mean values and 95% CIs are computed over all tested data, respectively. The highest scores are bold faced. "*" means the compared method has statistically significant differences ($p$<0.05) compared to our CardioMM under Wilcoxon signed-rank test.

Patient 1

Fully sampled
Radiologists’ score: 3.5

CardioMM (8x acceleration)
Radiologists’ score: 4.0

Disease: Arrhythmia + heart failure + pericarditis
Center: SHQT | Modality: LGE | Scanner: Siemens 1.5T Vida

Patient 2

Fully sampled
Radiologists’ score: 3.0

CardioMM (8x acceleration)
Radiologists’ score: 5.0

Disease: Dilated cardiomyopathy + heart failure + pericarditis
Center: SHQT | Modality: LGE | Scanner: UIH 1.5T umr680

**Supplementary Fig. 13 | Faster is more than speed—it represents improved image quality for challenging cases.** Long fully sampled scans sometimes break down in severe CVD patients due to breath-hold difficulty and unreliable ECG gating, leading to severe motion artifacts, degraded image quality, and compromised clinical diagnosis. Ultra-fast acquisition with CardioMM reconstruction enables patient-friendly imaging and diagnostically reliable image quality for clinically challenging cases, since acceleration shortens the vulnerable time window where motion and ECG instability corrupt k-space data.

## Supplementary Note 9. Results of ablation study

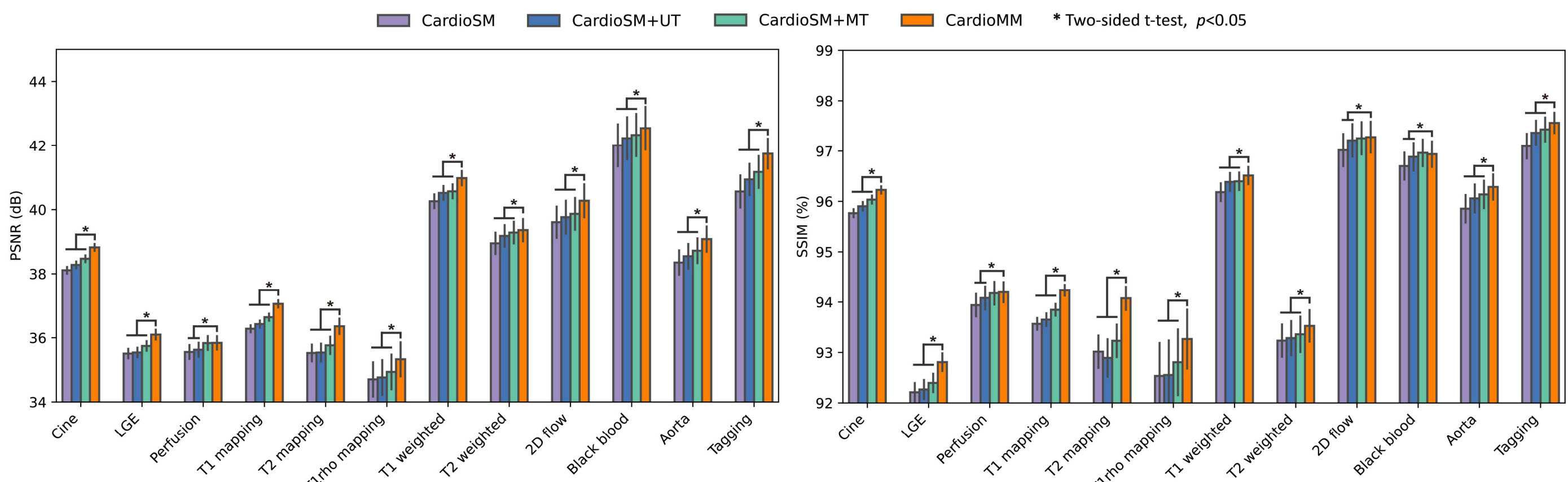

**Supplementary Fig. 14 | Ablation study.** Quantitative comparisons of reconstructions are shown for each modality, including PSNR and SSIM. Note: This evaluation is conducted across eight internal centers, using three undersampling patterns (uniform, random, radial) with varying AFs (8×–24×). The reported mean values and 95% CIs in the bar charts are computed over all tested data for each modality, respectively.

**Supplementary Table 10 | Ablation study across eight internal centers, using three undersampling patterns (uniform, random, radial) with varying AFs (8×–24×) [Mean (95% CI)].**

| Method | PSNR (dB) | SSIM (%) |
|---|---|---|
| CardioSM | 37.26 (37.17–37.34) * | 94.27 (94.19–94.35) * |
| CardioSM+UT | 37.39 (37.30–37.48) * | 94.35 (94.27–94.44) * |
| CardioSM+MT | 37.57 (37.49–37.66) * | 94.50 (94.42–94.59) * |
| CardioMM | **37.94 (37.86–38.03)** | **94.83 (94.76–94.90)** |

Note: This assessment involves 75,753 multi-coil k-space data from 1,495 scans of 320 participants, covering 12 CMR modalities acquired on routine high-field scanners (1.5T and 3.0T). The mean values and 95% CIs are computed over all tested data, respectively. The highest PSNR and SSIM values are bold faced. “*” means the compared method has statistically significant differences ($p$<0.05) compared to our CardioMM under two-sided t-test.